\documentclass[sigplan, nonacm]{acmart}
\usepackage{color}
\usepackage[normalem]{ulem}
\usepackage{xspace}
\usepackage{subcaption}
\usepackage{booktabs}
\usepackage{enumitem}
\usepackage{algorithm}
\usepackage{algpseudocode}

\newcommand{\sysn}{ThermE}
\newcommand{\C}{$^\circ$C\xspace}
\begin{document}
\title[ThermE]{ThermE: Predictive Management of Shared Thermal Headroom for Sustained LLM Inference on Thermally Constrained Edge SoCs}

\author{Rui Lu}
\affiliation{
  \institution{The Hong Kong Polytechnic University}
  \city{Hong Kong}
  \country{China}
}
\email{rui2020.lu@connect.polyu.hk}

\author{Yuheng Wang}
\affiliation{
  \institution{Southern University of Science and Technology}
  \city{Shenzhen}
  \country{China}
}
\email{12312317@mail.sustech.edu.cn}

\author{Bozheng Liu}
\affiliation{
  \institution{Southern University of Science and Technology}
  \city{Shenzhen}
  \country{China}
}
\email{12413116@mail.sustech.edu.cn}

\renewcommand{\shortauthors}{Lu et al.}

  \begin{abstract}

Compact edge system-on-chip (SoC) platforms increasingly run sustained LLM inference under thermal constraints, while their CPU, GPU, and RAM share a cooling path. Prefill and decode therefore consume shared, time-varying thermal headroom, yet vendor governors react only near hardware throttling thresholds without knowledge of request state or upcoming work. A control decision that improves current performance can thus consume headroom too quickly and degrade subsequent service.
In this paper, we present \textit{ThermE}, a runtime system that predicts and jointly manages shared thermal headroom for sustained LLM inference on edge SoCs. Its Fast LLM-to-Heat Compiler maps the model, requests, runtime state, and candidate actions to domain heats without executing LLMs. A partial differential equation (PDE)-constrained Headroom Predictor uses ThermPINN for offline thermal identification and a Reduced Headroom Predictor (RHP) for low-overhead online uncertainty-calibrated headroom forecasts. An Uncertainty-Aware Action Scheduler then selects actions that balance serving quality and future headroom. We implement ThermE atop vLLM and evaluate it across four LLM inference workloads. The results show that ThermE reduces TTFT and TPOT by 40.55\% and 12.17\%, respectively, relative to vLLM. It achieves a 5.70\% SLO violation rate, compared with 12.30\% for the strongest baseline, while its predictor obtains a 1.94$^\circ$C MAE with 11.50 ms overhead.

  \end{abstract}

  \maketitle

\section{Introduction}
\label{sec-introduction}

\begin{figure}[t]
  \centering
  
  \includegraphics[width=0.85\columnwidth]{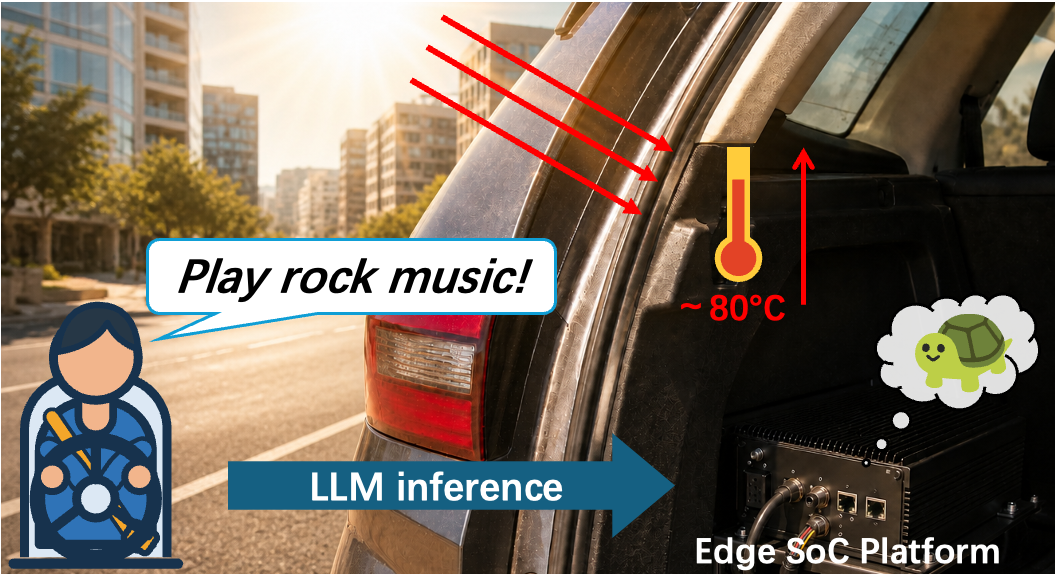}
  \caption{A thermally exposed edge platform in a solar-heated truck reduces the headroom and degrades requests.}
  \label{fig:therm_scense}
  
\end{figure}

Large language model inference is moving onto compact edge system-on-chip (SoC) platforms deployed in thermally constrained environments, such as satellites and industrial robots.
More broadly, global warming and increasingly frequent heat waves expose otherwise conventional edge computers in roadside cameras, robots, drones, etc to high temperatures~\cite{ipcc-ar6-wg1-spm,epa-extreme-heat}.
For example, edge computers on autonomous vehicles are often placed in compact trunks, where solar loading can raise internal temperatures well above ambient, i.e, $61$\C to $80$\C in direct sunlight~\cite{horak2017cabin}, as shown in Fig.~\ref{fig:therm_scense}.
These conditions raise the device's ambient temperature and reduce \textit{thermal headroom}, which is defined as the remaining temperature margin before thermal protection constrains execution.

\begin{figure}[t]
  \centering
  
  \includegraphics[width=\columnwidth]{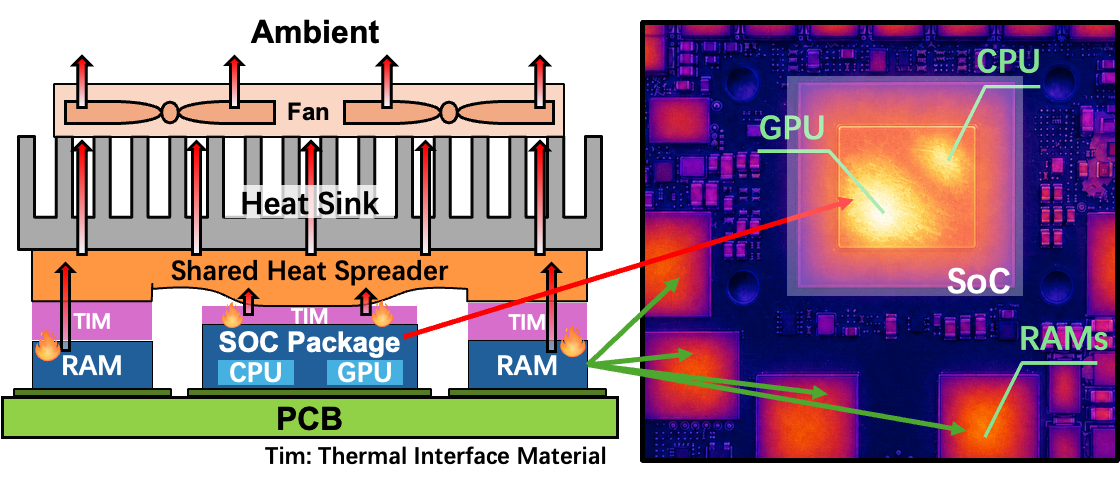}
  \caption{Cross-sectional schematic of the Nvidia Jetson AGX Orin thermal stack~\cite{nvidia-agx-orin-tdg}. Its CPU, GPU, and RAM share a heat spreader, heat sink, and fan-assisted cooling path, so activity in one domain reduces thermal headroom for all.}
  \label{fig:thermal_coupling}
  
\end{figure}

Meanwhile, semiconductor scaling concentrates increasingly demanding LLM computation within compact edge SoC modules, where sustained workloads continuously generate heat.
Unlike server platforms, where CPUs, GPUs, and RAM are discrete devices with largely independent power and thermal controls, compact edge SoCs couple these domains through the thermal stack shown in Fig.~\ref{fig:thermal_coupling}. Heat crosses their thermal interface materials before converging at a shared heat spreader, heat sink, and fan-assisted cooling path.
Heat generated by one domain therefore reduces the headroom available to others. Because the shared cooling path dissipates heat more slowly than inference workloads change, this shared headroom varies over time. Deployment conditions determine its initial value, while workload and cooling conditions determine its subsequent availability.

Hardware thermal protection reacts as shared headroom narrows, exposing two limitations for sustained LLM inference on thermally constrained edge SoCs.
First, vendor governors are reactive and application-independent. They activate protection from measured temperature, current, or power, instead of the LLM phase, request state, remaining tokens, or upcoming work.
As temperatures approach throttling thresholds, the platform would reduce frequencies by up to 50\%~\cite{nvidia-jetson-orin-thermal}, degrading sustained service.
Second, existing inference controls lack visibility into their effects on future thermal headroom. Fixed DVFS settings, power caps, batching, and offloading adjust instantaneous power or latency, but do not account for future thermal headroom.
A performance-improving control decision may consume shared headroom too quickly and trigger lower frequencies later.

To understand these limitations, we conduct three motivational experiments that reveal three properties of sustained LLM inference under thermal constraints on edge SoCs.
First, built-in thermal throttling reacts only after thermal headroom has been largely consumed and then degrades serving quality through staged frequency reductions.
Second, the CPU, GPU, and SoC thermal zones heat and cool together through the shared package.
Third, hardware profiles with different CPU/GPU frequencies trade off time to first token (TTFT), time per output token (TPOT), and headroom consumption. Together, these observations show that thermal headroom is a shared, time-varying resource that must be managed proactively according to serving demand and thermal state.

These observations expose three critical challenges.
First, the heat generated by the upcoming workload must be estimated before execution for each candidate control action.
Second, current temperatures alone do not reveal how the heat generated by each candidate action will affect future headroom. Heat propagates across coupled domains before dissipating through the shared cooling path, and this process depends on retained heat from recent activity, ambient temperature, and cooling state.
Third, preserving headroom competes with serving performance. Hardware frequencies and scheduled work jointly determine latency and how much thermal headroom remains for later requests. The runtime must select actions that preserve serving quality under thermal constraints while accounting for uncertain future temperatures and the risk of hardware throttling.

In this paper, we present \textbf{\textit{\sysn{}}}, a runtime system that predictively manages shared thermal headroom for sustained LLM inference on thermally constrained edge SoCs.

A \textit{Fast LLM-to-Heat Compiler} maps model and request descriptors, runtime state, and candidate actions to CPU, GPU, and RAM heat inputs without executing the LLM.
A \textit{Partial Differential Equation (PDE)-Constrained Headroom Predictor} uses ThermPINN to identify coupled thermal dynamics offline and construct reduced thermal operators.
Online, the Reduced Headroom Predictor (RHP) uses these operators to produce uncertainty-calibrated headroom estimates with low computational overhead.
An \textit{Uncertainty-Aware Action Scheduler} retains the current action when service quality and thermal runway are sufficient.
Otherwise, beam search selects a new action. The scheduler applies this action and updates its decision after the next observation.

We implemented \sysn{} atop vLLM on an Nvidia Jetson AGX Orin. Under the Stress workload at the 65\C ambient setpoint, \sysn{} reduced TTFT, TPOT, and p95 end-to-end latency by 40.55\%, 12.17\%, and 13.01\%, respectively, relative to vLLM, and achieved a 5.70\% SLO violation rate compared with 12.30\% for the strongest baseline. The headroom prediction obtained a 1.94\C MAE with $24.36\,\mathrm{ms}$ overhead.
In summary, this paper makes the following contributions.
\begin{itemize}[leftmargin=*,itemsep=0pt,topsep=2pt]
  \item We characterize sustained LLM inference under thermal constraints on an edge SoC and show that cross-domain thermal coupling, reactive hardware throttling, and the serving-performance tradeoffs of hardware profiles make thermal headroom a shared, time-varying runtime resource.
  \item We design \sysn{} to compile candidate workloads into domain heat, combine offline physics-constrained thermal identification with low-overhead online headroom prediction, and coordinate hardware profiles and scheduled-token caps under hard temperature constraints.
  \item We implement \sysn{} on an Nvidia Jetson AGX Orin and evaluate its serving quality, thermal behavior, prediction accuracy, sensitivity, component contributions, and runtime overhead across representative LLM workloads and thermal conditions.
\end{itemize}

\section{Background}
\label{sec:background}

\begin{figure}[t]
  \centering
  
  \includegraphics[width=\columnwidth]{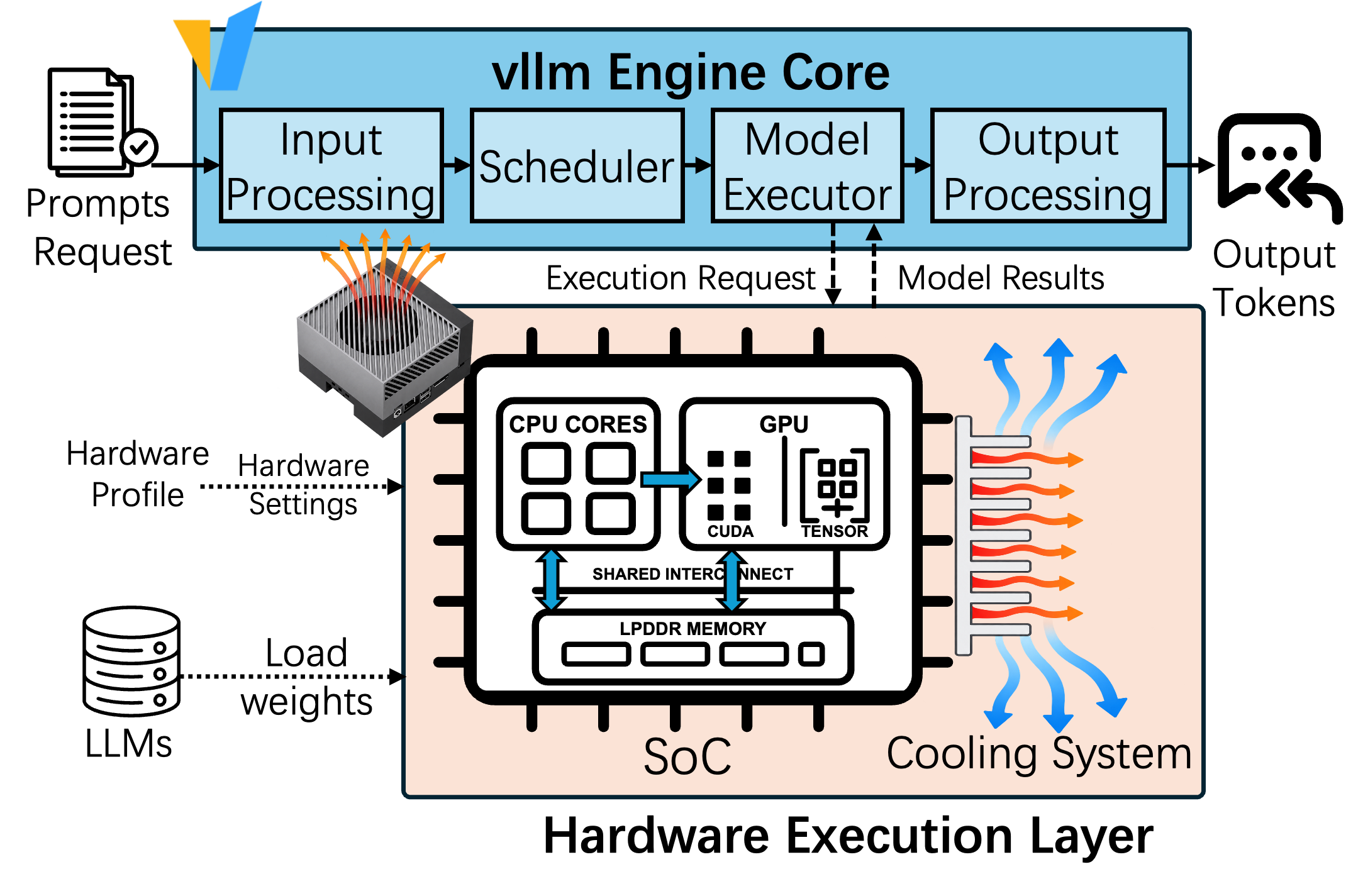}
  \caption{LLM inference execution on edge SoCs. vLLM maps request processing, scheduling, model execution, and output processing to CPU, GPU, and shared RAM. }
  \label{fig:vllm-jetson-workflow}
  
\end{figure}

\subsection{Thermal Characteristics of Edge SoC Platforms}
A system-on-chip (SoC) integrates multiple computing functions on a single chip, and similarly, a system-on-package (SoP) integrates multiple chips or SoCs within a compact package.
These platforms place CPUs, GPUs, memory controllers, and memory within a compact module. Recent inference systems exploit these heterogeneous resources and their tightly integrated memory paths \cite{xu2025fastnpu,jung2025aria,moon2025hybe,li2025h2llm}. Their compute and memory domains commonly share a heat spreader, heat sink, and fan, which couple their thermal behavior through a common cooling path.

Thermal headroom is the remaining temperature margin between the current thermal state and the hardware limit. Ambient temperature and the cooling state determine the initial margin. During execution, dynamic power generates heat, while the package, heat sink, enclosure, and surrounding air dissipate heat. Because these structures dissipate heat more slowly than hardware activity changes, the current headroom reflects both the present workload and heat retained from recent activity \cite{stojkovic2025tapas}.

\subsection{LLM Inference Execution on Edges}
\label{sec:background-vllm}

LLM inference consists of a \emph{prefill} phase that processes the input prompt and constructs the key-value cache, followed by a \emph{decode} phase that generates one token at a time from the model state and growing cache \cite{kamath2025podattention,feng2025windserve,stojkovic2025tapas,cao2025amali,lee2025aif,moon2025hybe}. Service quality depends on both phases, with prefill largely determining TTFT and decode characterized by TPOT and sustained output throughput \cite{kamath2025podattention,feng2025windserve}.

vLLM is a widely adopted open-source LLM serving engine that combines request scheduling, batching, KV-cache management, and model execution~\cite{kwon2023vllm}, covering half of the text inference services~\cite{runpod2026stateofai}.
Fig.~\ref{fig:vllm-jetson-workflow} shows how vLLM maps this execution onto an edge SoC. Input Processing prepares requests, the Scheduler allocates a token budget and forms continuous batches, the Model Executor dispatches kernels, and Output Processing returns generated tokens \cite{vllm-v1-guide}. The CPU handles request processing, scheduling, and kernel launches. The integrated GPU executes model kernels, while shared LPDDR RAM stores model weights and the KV cache.
On Nvidia Jetson, the vLLM platform layer adapts this loop to the ARM CPU and integrated GPU memory architecture \cite{nvidia-jetson-vllm}. The scheduled work and active hardware profile together determine the CPU, GPU, and memory activity of each inference interval.

\subsection{Hardware Thermal Headroom Management}
Edge SoCs expose multiple thermal zones whose sensors track the temperatures of the CPU, GPU, memory, and package. Each zone has configured trip points for progressively stronger protection. Nvidia Jetson platforms, for example, reduce CPU and GPU frequencies as measured temperatures cross their throttling thresholds \cite{nvidia-jetson-orin-thermal}. Firmware and kernel drivers may also adjust voltage, power limits, cooling state, or task placement.
If high temperatures persist despite these controls, a thermal zone can eventually reach its critical trip point and trigger device shutdown \cite{nvidia-jetson-orin-thermal}.

These mechanisms manage thermal headroom through a feedback loop. Sensors report the current thermal state, the hardware governor compares each zone against its trip points, and actuators adjust power or cooling. The available platform headroom follows the current zone temperatures and their configured limits. This hardware loop operates continuously beneath the inference framework and can change the effective CPU and GPU frequencies as temperatures evolve.
Section~\ref{sec:characterization} measures how this hardware-managed headroom interacts with sustained LLM inference.

\section{Motivation}
\label{sec:characterization}
In this section, we characterize how sustained LLM inference interacts with shared thermal headroom on edge SoCs through three experiments examining hardware throttling, cross-domain thermal coupling, and the serving--thermal tradeoffs of different hardware profiles.
Our experiments investigate three questions:
\begin{itemize}
  \item \textit{When} and \textit{how} do the built-in thermal throttling mechanisms influence LLM inference?
  \item \textit{How} does activity in one hardware domain affect the thermal state of other zones?
  \item \textit{How} do frequency setups trade off TTFT, TPOT, and thermal headroom under different thermal conditions?
\end{itemize}

\begin{figure}[t]
  \centering

  \begin{subfigure}[t]{\columnwidth}
    \centering
    \includegraphics[width=\linewidth]{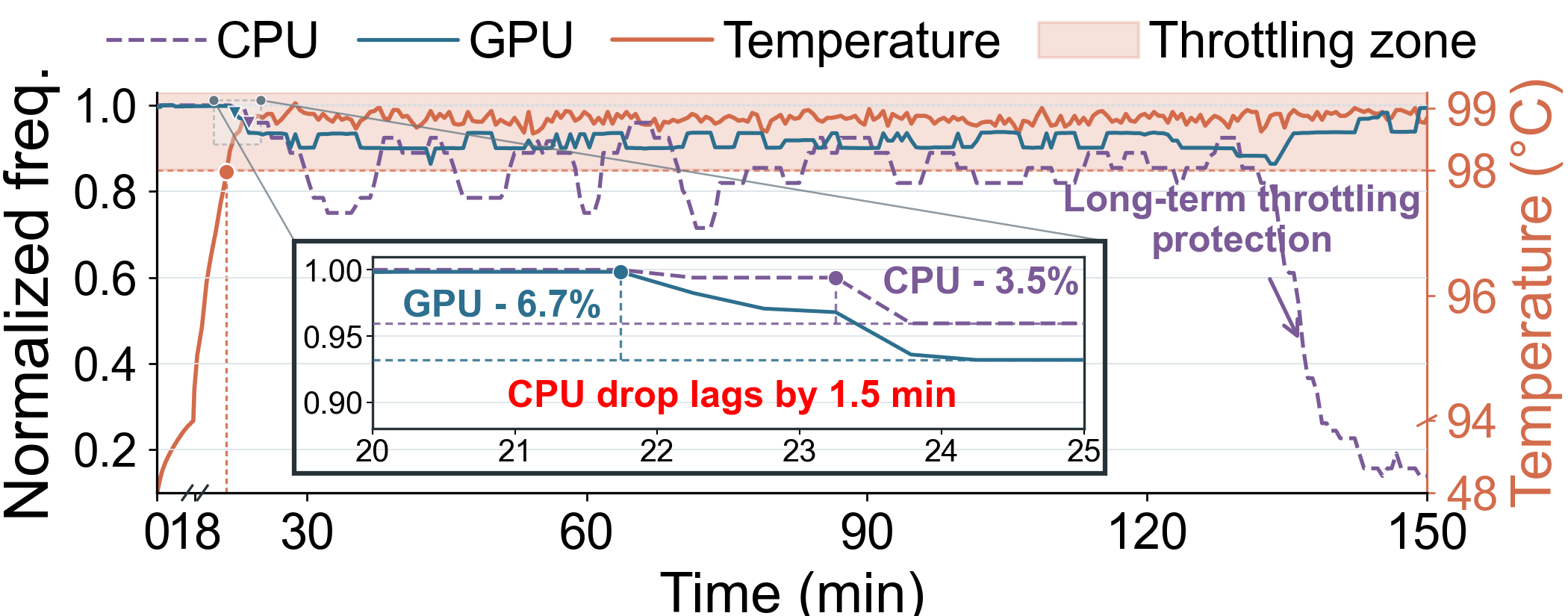}
    \caption{Thermal trajectory and normalized CPU and GPU frequencies.}
    \label{fig:motivation-thermal-a}
  \end{subfigure}
  \begin{subfigure}[t]{\columnwidth}
    \centering
    \includegraphics[width=\linewidth]{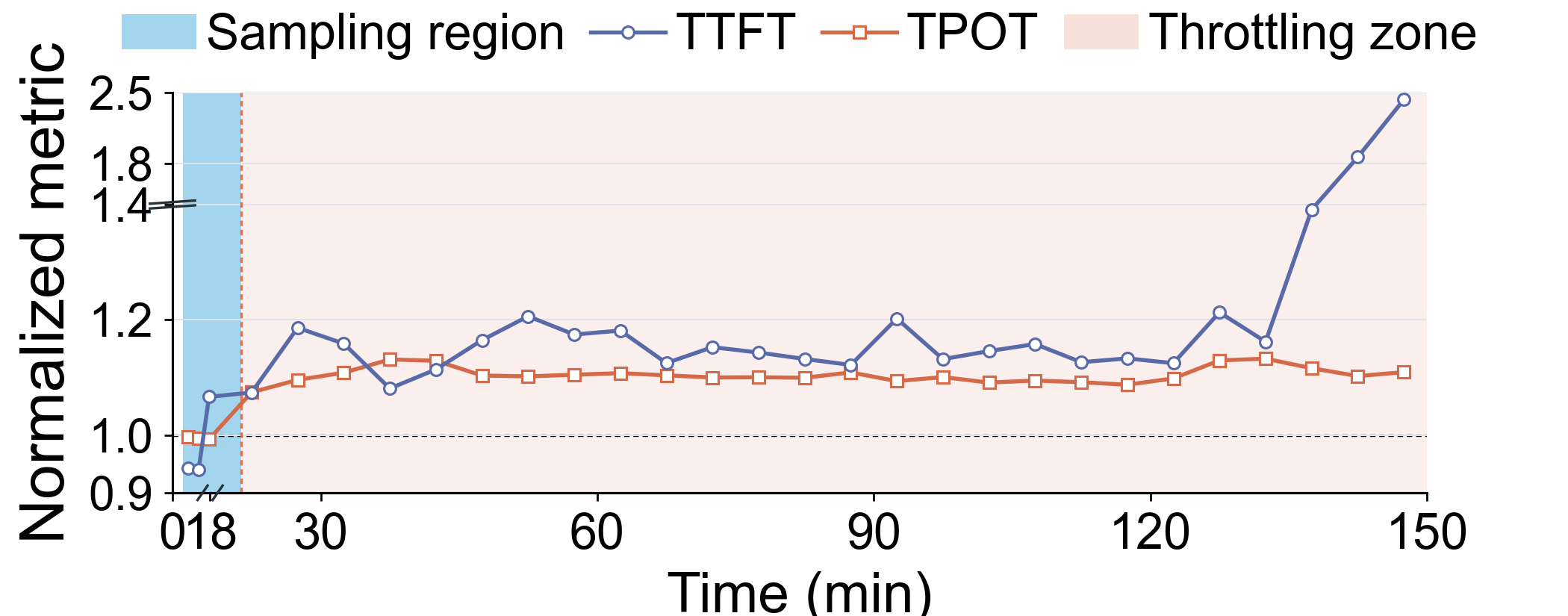}
    \caption{Serving quality degradation during throttling.}
    \label{fig:motivation-thermal-b}
  \end{subfigure}
  \caption{Built-in thermal control reacts after the device enters the near-limit region and progressively degrades LLM inference performance.}
  \label{fig:motivation-thermal}
  
\end{figure}

\noindent\textbf{Setups.}
We deploy the experiments on an Nvidia Jetson AGX Orin 64GB.
We place it in an electric thermostatic chamber that can control the temperature ranging from 48\C to 65\C.
We collect thermal-zone temperatures, CPU and GPU frequencies, power, TTFT, and TPOT as applicable.

\noindent\textbf{Observation 1: Built-In Thermal Throttling Is Reactive and Preset.}
We first examine how the built-in thermal mechanism responds during sustained LLM inference on \texttt{Qwen3:8B} at 48.0\C.
Fig.~\ref{fig:motivation-thermal}\subref{fig:motivation-thermal-a} shows the maximum SoC temperature and normalized CPU and GPU frequencies over a 150-minute continuous run. To observe details of hardware response near the throttling point, we enlarge a five-minute window inside. Fig.~\ref{fig:motivation-thermal}\subref{fig:motivation-thermal-b} records the normalized TTFT and TPOT before throttling.

The results show both CPU and GPU initially remain at baseline frequencies while the maximum temperature rises to the 98\C near-limit region.
The first sustained GPU frequency reduction begins at 22.25 minutes and 98.56\C, with GPU frequency falling by 6.7\%, followed 1.53 minutes later by a 3.5\% reduction in CPU frequency.
The full trace then shows additional staged reductions, followed by persistent CPU throttling that further lowers the CPU frequency. The frequency recovers only after the testbed is restarted.

From Fig.~\ref{fig:motivation-thermal}\subref{fig:motivation-thermal-b}, the serving performance metrics degrade during throttling.
During the first 20--25 minutes, which contain the first sustained frequency reductions, TTFT and TPOT both reach $1.07\times$ their respective sampling-region means. After prolonged throttling, TTFT reaches $2.43\times$ from 45--150 minutes, while TPOT reaches $1.11\times$.
The results show that the default throttling mechanism activates after most thermal headroom has been consumed and reduces serving performance through staged clock reductions.

\noindent\textit{\underline{Insight \#1.}} Built-in throttling reduces serving performance after thermal headroom has largely been consumed. This motivates control decisions that account for future headroom before hardware protection intervenes.

\begin{figure}[t]
  \centering
  
  \includegraphics[width=\columnwidth]{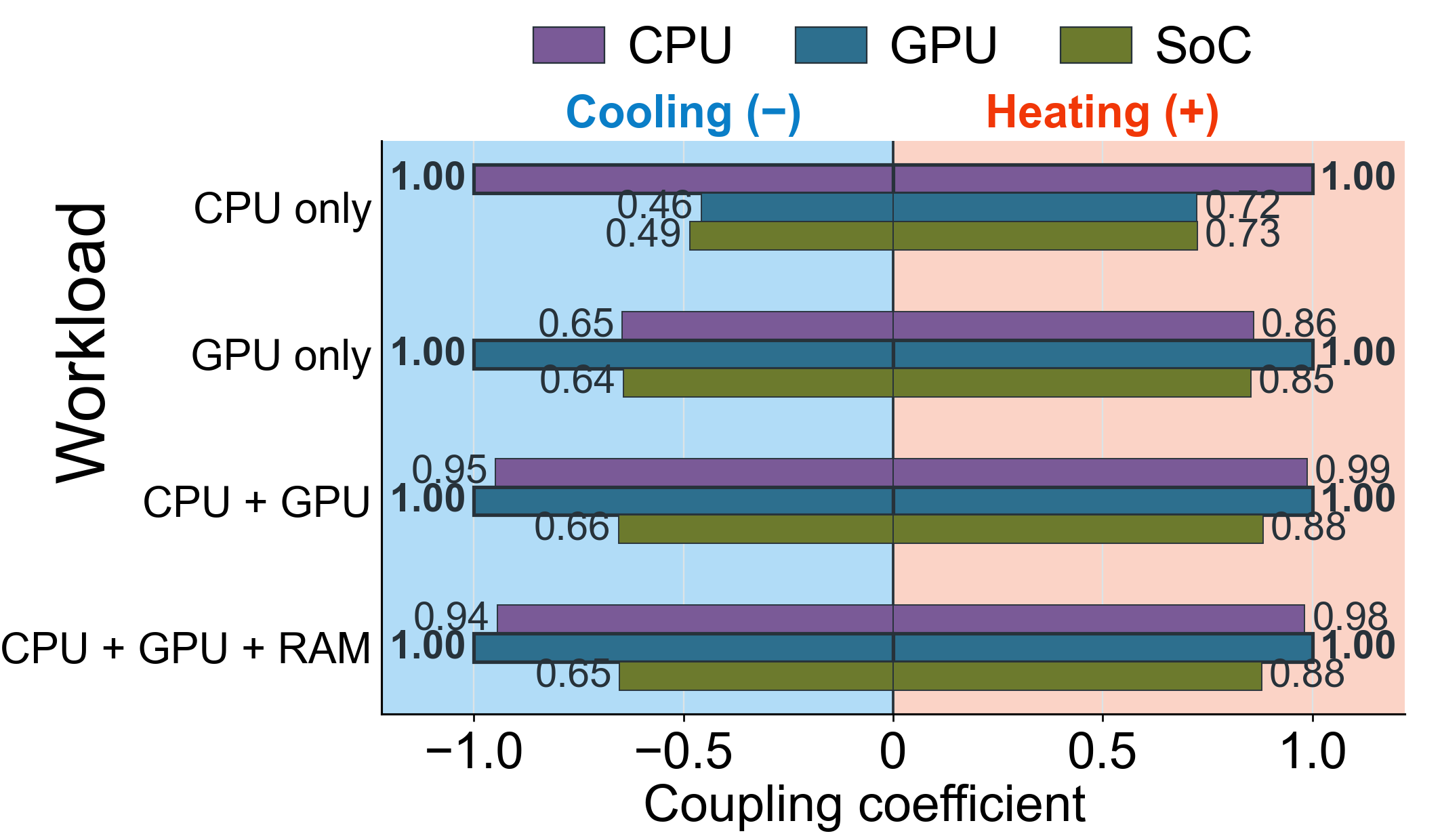}
  \caption{Thermal coupling among CPU, GPU, and SoC zones under four workloads. Each coefficient reports the observed-zone temperature change per 1\C change in the reference zone during heating ($+$) or cooling ($-$).}
  \label{fig:shared-thermal-headroom}
  
\end{figure}

\noindent\textbf{Observation 2: Shared Cooling Couples SoC Thermal Domains.}
We next quantify thermal coupling among the CPU, GPU, and SoC thermal zones under \texttt{CPU-only}, \texttt{GPU-only}, \texttt{CPU+GPU}, and \texttt{CPU+GPU+RAM} workloads.
We stress the CPU, GPU, and RAM using an arithmetic loop, repeated $2048\!\times\!2048$ CUDA matrix multiplications, and four workers continuously copying 256-MiB byte arrays, respectively.
Fig.~\ref{fig:shared-thermal-headroom} reports steady temperature-rise ratios during heating and fitted temperature-drop ratios after each workload stops, normalized by the reference zone coefficient.

Under the \texttt{CPU-only} heating process, each 1\C CPU rise results in 0.72\C and 0.73\C rises in the GPU and SoC zones. Under \texttt{GPU-only} heating, each 1\C GPU rise results in 0.86\C and 0.85\C rises in the CPU and SoC zones.
The concurrent workloads produce similar CPU and GPU temperature rises.
Relative to the GPU reference, the CPU heating coefficient is 0.99 for \texttt{CPU+GPU} and 0.98 for \texttt{CPU+GPU+RAM}, while the corresponding SoC coefficient is 0.88 in both cases.
During cooling, the off-reference coefficients remain between 0.46 and 0.95, which shows that the zones also recover together after activity stops.

These off-reference coefficients arise as the zones share the package and cooling path. Heat generated by one domain propagates via the common heat spreader/sink, and recovery in one zone remains linked to the stored heat of the others. A control decision for one domain therefore changes the thermal state and remaining headroom of the whole package.

\noindent\textit{\underline{Insight \#2.}} CPU, GPU, SoC thermal zones heat and cool together. Predicting future headroom therefore requires accounting for cross-domain heat propagation and retention.

\begin{figure}[t]
  \centering
  
  \begin{subfigure}[t]{0.485\columnwidth}
    \centering
    \includegraphics[width=\linewidth]{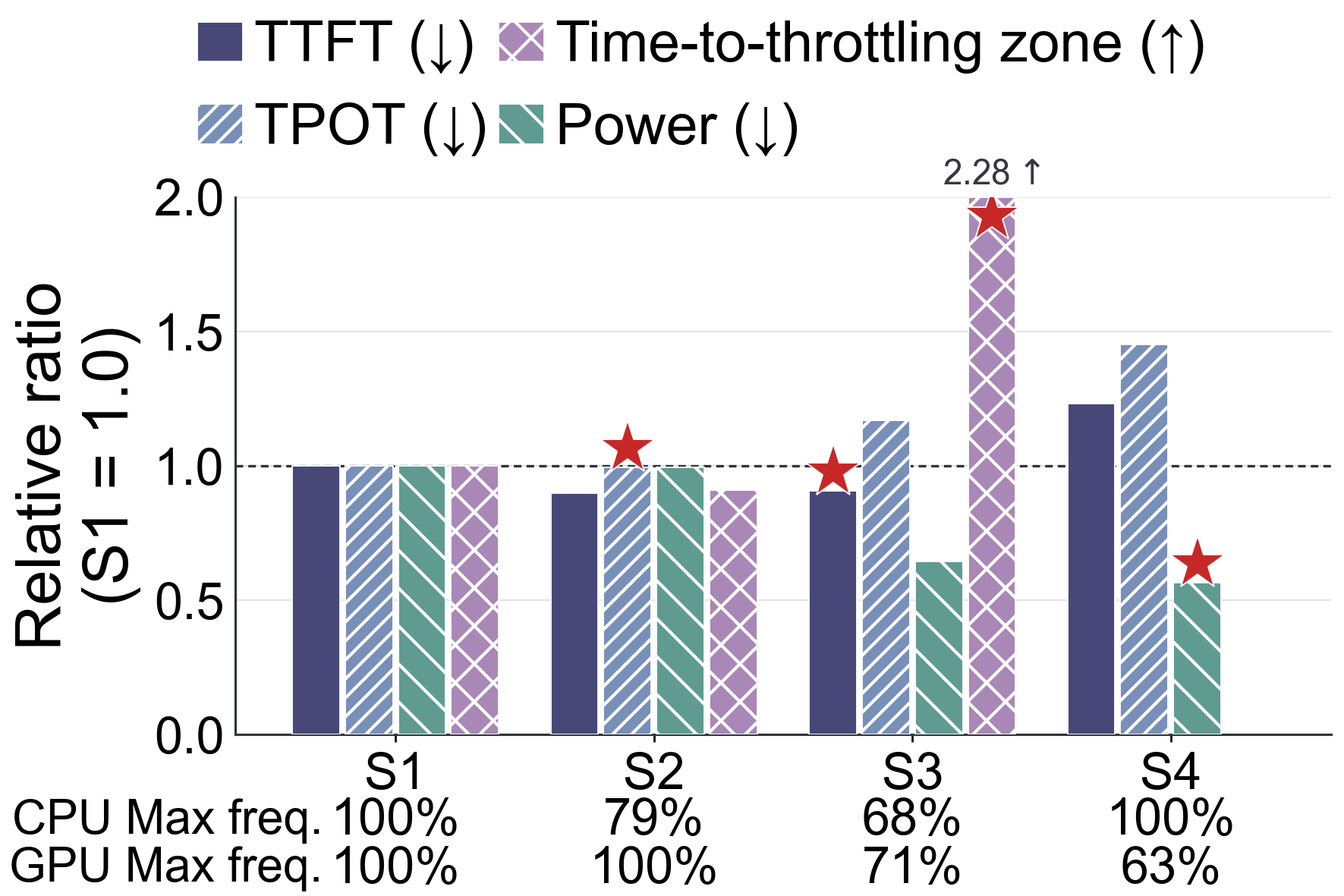}
    \caption{Before the throttling zone.}
    \label{fig:static-runtime-phase-tradeoff-a}
  \end{subfigure}\hfill
  \begin{subfigure}[t]{0.485\columnwidth}
    \centering
    \includegraphics[width=\linewidth]{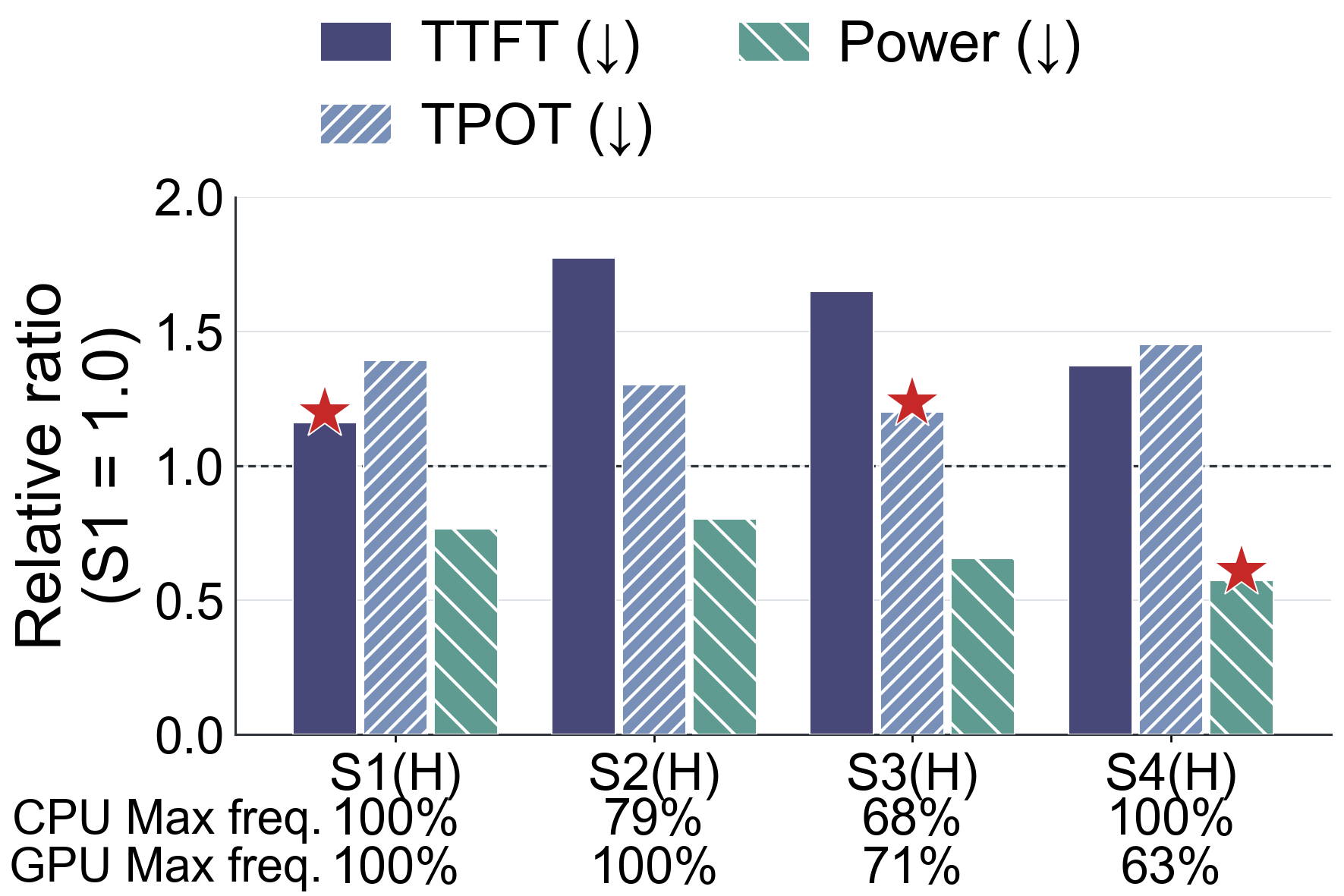}
    \caption{Within the throttling zone.}
    \label{fig:static-runtime-phase-tradeoff-b}
  \end{subfigure}
  \caption{Serving, power, and thermal-headroom tradeoffs before and within the throttling zone.}
  \label{fig:static-runtime-phase-tradeoff}
  
\end{figure}

\noindent\textbf{Observation 3: Hardware Profiles Trade Off Serving Performance and Thermal Headroom.}
Fig.~\ref{fig:static-runtime-phase-tradeoff} compares four hardware profiles with different CPU/GPU frequencies before and within the throttling zone, noted as (H). All metrics are normalized to S1. $t_{\mathrm{zone}}$ measures time to entry into the throttling zone.

The results show that S2 reduces TTFT by 10.13\%. S3 reduces power by 35.57\% and increases $t_{\mathrm{zone}}$ by 128.46\%, but increases TPOT by 16.97\%. S3(H) and S1(H) increase TPOT by 19.97\% and 39.38\%, and TTFT by 65.01\% and 16.19\%, respectively. Thus, S3(H) achieves lower TPOT than S1(H) at the cost of higher TTFT.
S4 and S4(H) reduce power by 43.37\% and 42.59\%, respectively. S4 increases TTFT and TPOT by 23.17\% and 45.26\%, while S4(H) increases them by 37.41\% and 45.30\%. These results show that lower power comes at the cost of longer TTFT and TPOT, so minimizing power alone does not preserve serving performance.

\noindent\textit{\underline{Insight \#3.}} Hardware profiles trade off TTFT, TPOT, and the rate of thermal-headroom consumption. Hardware profile selection must balance TTFT and TPOT while accounting for serving demand and remaining thermal headroom.

Overall, these insights suggest that sustained LLM serving requires proactive, coordinated hardware control that anticipates shared thermal headroom and balances service quality under changing workload and thermal conditions.

\section{Design of ThermE}
\label{sec:system}

\subsection{Design Overview}
\label{sec:design-overview}

\sysn{} is a shared-headroom management system for sustained LLM inference on shared-cooling edge SoC platforms atop the vLLM serving engine. At each control interval, it selects a control action that pairs a hardware profile with a scheduled-token cap to balance serving performance and heat generation under limited shared thermal headroom.

\begin{figure}[t]
  \centering
  
  \includegraphics[width=\linewidth]{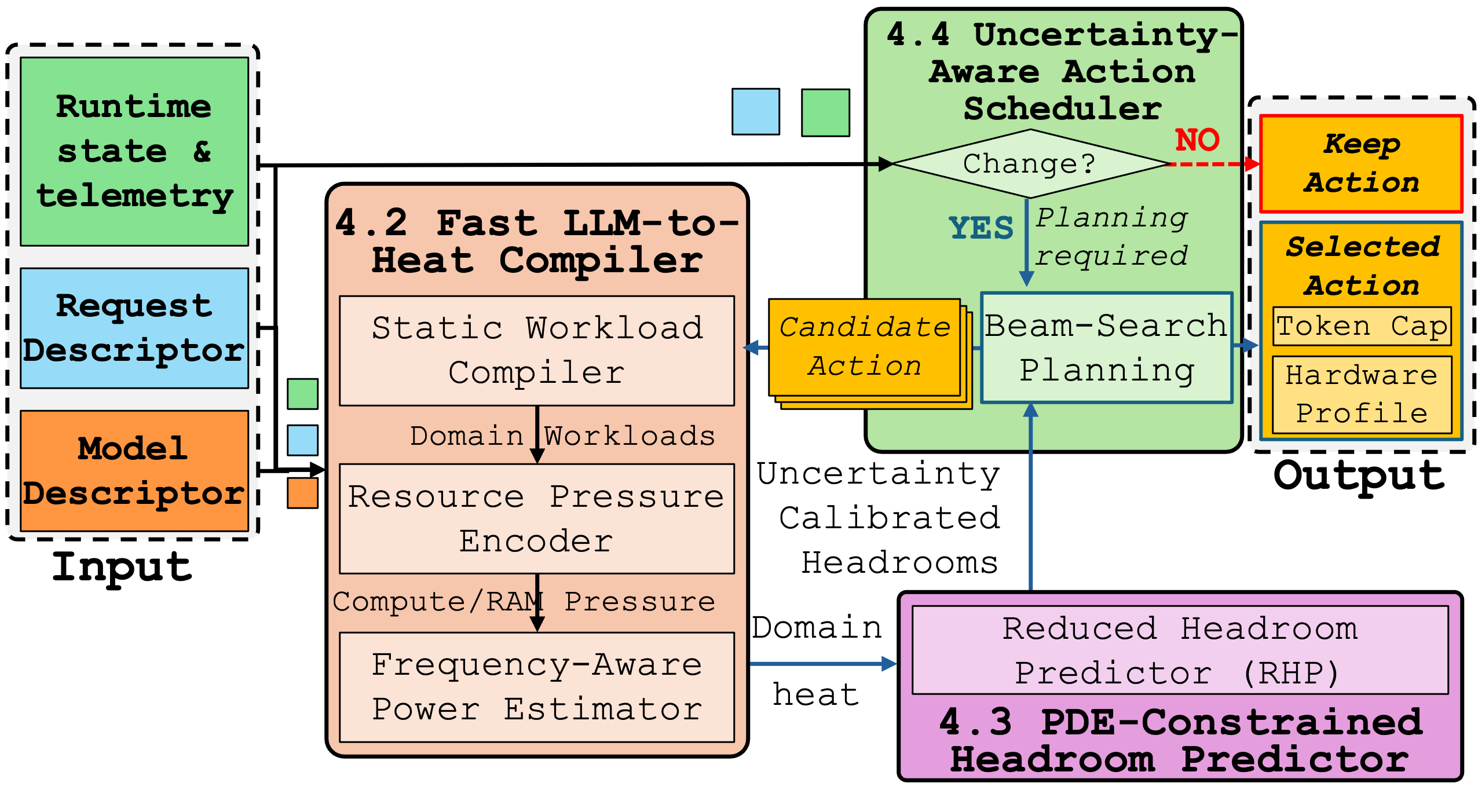}
  \caption{System architecture of \sysn{}. The compiler estimates domain heat inputs for candidate actions, the predictor provides uncertainty-calibrated headrooms, and the scheduler selects actions that pair hardware profiles with scheduled-token caps.}
  \Description{System architecture with model and request descriptors, runtime state and telemetry, a fast LLM-to-heat compiler, a PDE-constrained headroom predictor, and an uncertainty-aware action scheduler. The scheduler retains the current action or selects a new action through beam search.}
  \label{fig:system-architecture}
  
\end{figure}

The three inputs in Fig.~\ref{fig:system-architecture} are the \textit{request descriptor}, \textit{runtime state and telemetry}, and \textit{model descriptor}. They specify request demand, current interval work and operating conditions, and model architecture and quantization configuration, respectively. Each action pairs one hardware profile with a scheduled-token cap. A hardware profile specifies settings for the controls exposed by the platform adapter. The cap bounds the work admitted to the next control interval.

\sysn{} combines three coupled components in Fig.~\ref{fig:system-architecture}.
\begin{itemize}[itemsep=0pt,parsep=0pt,topsep=0pt,itemindent=0pt,labelsep=5pt,leftmargin=10pt]
  \item \emph{Fast LLM-to-Heat Compiler.} The compiler maps the model and request descriptors, runtime state, candidate actions to CPU, GPU, RAM heat inputs without executing the LLM.
  \item \emph{PDE-Constrained Headroom Predictor.} ThermPINN identifies coupled thermal dynamics offline. Online, the RHP predicts temperatures from telemetry and candidate heat inputs, then computes uncertainty-calibrated per-zone and shared headrooms.
  \item \emph{Uncertainty-Aware Action Scheduler.} Beam search selects hard-safe sequences, balancing service and energy costs with thermal runway during normal operation and the cumulative shared-headroom deficit during headroom-recovery operation.
\end{itemize}

\sysn{} separates the inference data path from the thermal control path. Its asynchronous controller evaluates candidate action sequences outside the inference critical path.

\subsection{Fast LLM-to-Heat Compiler}
\label{sec:llm-to-heat-compiler}

The \textit{Static Workload Compiler} extracts compute and RAM workloads that can be reused across candidate hardware profiles. The \textit{Resource Pressure Encoder} encodes these workloads as resource demands, which the \textit{Frequency-Aware Power Estimator} combines with candidate hardware profiles to estimate power. This separation limits candidate evaluation to the lightweight encoder and estimator.

\subsubsection{Static Workload Compiler}
This workload compiler obtains compute and RAM workloads from the request descriptor and model descriptor, which encodes the model architecture and quantization configuration.

The request descriptor $\mathbf{d}_{\mathrm{req}}=(S_{\mathrm{in}},O_{\mathrm{out}},C_{\mathrm{req}})$ specifies the input length, output-token budget, and number of concurrent requests represented by the compiled workload, respectively.
The compiler represents four types of static workload. MMA workloads represent Tensor Core work across the projection GEMMs in each Transformer layer~\cite{nvidia-cutlass-gemm}. Non-MMA GPU work covers attention data movement, element-wise operations, normalization, and other CUDA kernels. CPU serving-runtime workloads cover the host-side request preparation, sequence management, and execution dispatching. RAM workloads capture traffic using the weight footprint $B_w$ and per-token KV-cache size $B_{\mathrm{KV}}$.
Each workload combines prefill and decode
\begin{equation}
  W_{x}^{\mathrm{req}}(O_{\mathrm{out}})
  =W_{x}^{\mathrm{pre}}+O_{\mathrm{out}}W_{x}^{\mathrm{dec}}.
\label{eq:request-mma-workload}
\end{equation}
where $x\in\{\mathrm{MMA},\mathrm{CUDA},\mathrm{CPU},\mathrm{RAM}\}$.
Here, $W_{x}^{\mathrm{pre}}$ denotes prefill work, and $W_{x}^{\mathrm{dec}}$ denotes work per decode step. The encoder also retains the separate MMA phase workloads.

The runtime state and telemetry provide the current prefill and decode work, while each candidate action's scheduled-token cap bounds the work admitted to the next control interval. These workloads quantify compute operations and memory traffic at different scales, motivating separate pressure representations for power estimation.

\subsubsection{Resource Pressure Encoder}
The Resource Pressure Encoder uses two branches to encode compute workloads and model geometry, and RAM workloads and memory context, respectively. Each branch applies standardization, a linear mapping, and a sigmoid activation.
The memory context is $\mathbf{d}_{\mathrm{RAM}}=(B_w,B_{\mathrm{KV}},A_{\mathrm{ints}},b,f_q,b_{\mathrm{KV}})$, where $A_{\mathrm{ints}}$ denotes arithmetic intensity and $\nu_G$ is GPU frequency.
It forms the compute and RAM feature vectors as
\begin{equation}
\resizebox{\columnwidth}{!}{$\displaystyle
\mathbf{s}_C=[W_{\mathrm{MMA}}^{\mathrm{req}},W_{\mathrm{MMA}}^{\mathrm{pre}},W_{\mathrm{MMA}}^{\mathrm{dec}},W_{\mathrm{CUDA}}^{\mathrm{req}},W_{\mathrm{CPU}}^{\mathrm{req}},P,L,H,r_I,A_{\mathrm{att}},r_{\mathrm{KV}},\nu_G].
$}\nonumber
\end{equation}
\begin{equation}
\resizebox{0.675\linewidth}{!}{$\displaystyle
\mathbf{s}_M=[W_{\mathrm{RAM}}^{\mathrm{req}},\mathbf{d}_{\mathrm{RAM}},P,L,H,r_I,A_{\mathrm{att}},r_{\mathrm{KV}}].
$}
\end{equation}
A calibrated resource pressure encoder $\mathcal{E}_{\theta_{\mathrm{enc}}}$ maps these feature vectors to compute and memory pressure vectors
\begin{equation}
  (\mathbf{u}_C,\mathbf{u}_M)
  =\mathcal{E}_{\theta_{\mathrm{enc}}}(\mathbf{s}_C,\mathbf{s}_M).
\end{equation}
The vectors $\mathbf{u}_C$ and $\mathbf{u}_M$ summarize workload-dependent resource demands for the following frequency-aware power estimator.

\begin{figure}[t]
  \centering
  
  \includegraphics[width=0.975\linewidth]{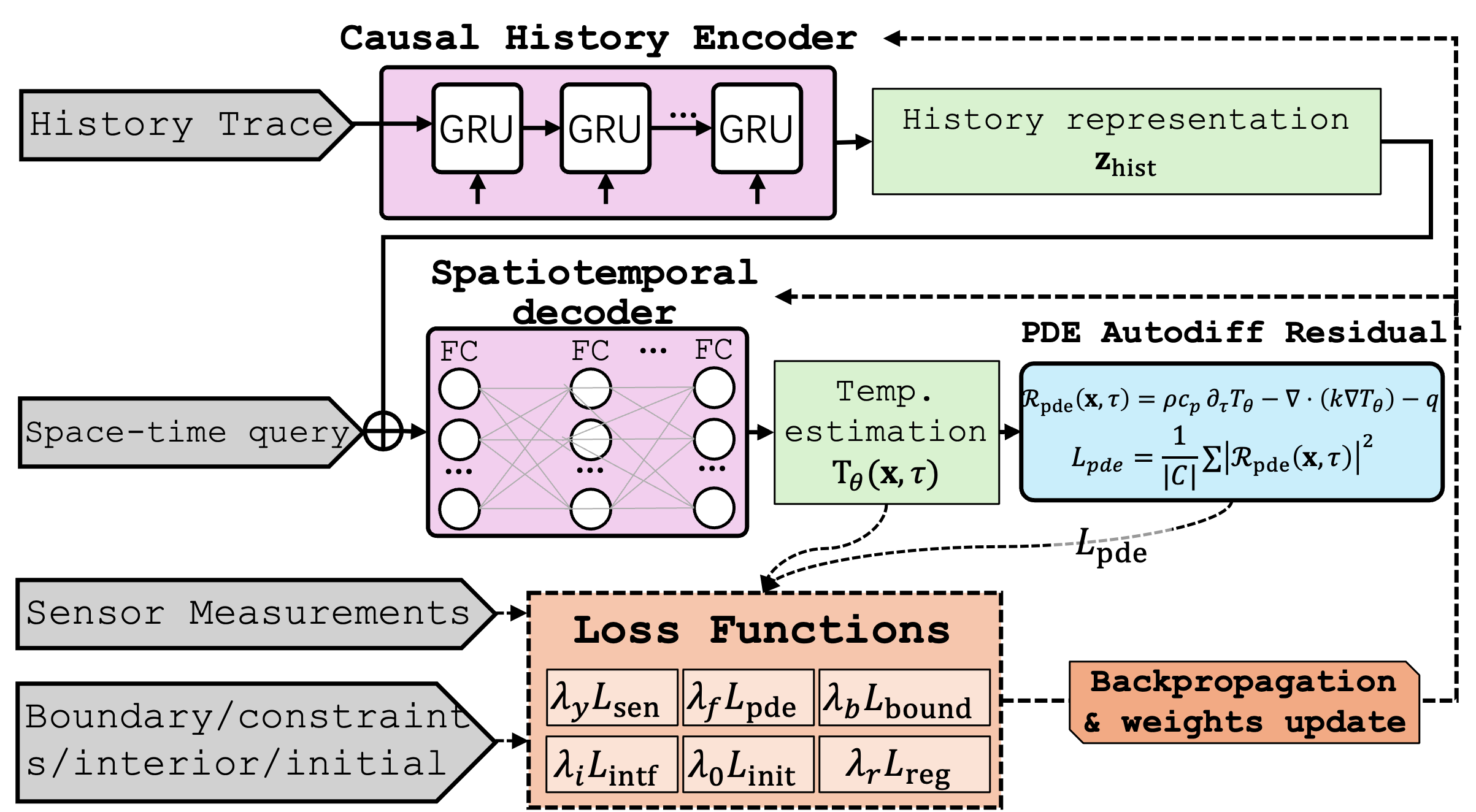}
  \caption{Offline ThermPINN training. The causal history encoder and spatiotemporal decoder estimate temperature under sensor and heat-transfer constraints.}
  \label{fig:thermpinn-training}
  
\end{figure}

\subsubsection{Frequency-Aware Power Estimator}
Let $\nu_C$ denote the standardized CPU frequency. Each validated CPU--GPU pair identifies one calibrated RAM frequency in the hardware profile. The estimator $\mathcal F_{\mathrm{power}}$ combines resource pressures and candidate frequencies to predict CPU, GPU, and RAM/SoC heat inputs
\begin{equation}
[p_{\mathrm{CPU}},p_{\mathrm{GPU}},p_{\mathrm{RAM}}]^\top
=\mathcal F_{\mathrm{power}}(\mathbf u_C,\mathbf u_M,\nu_G,\nu_C).
\end{equation}
This mapping incorporates matched-idle subtraction and total-power-constrained source allocation. The estimator exposes these three domain heat inputs directly to the headroom predictor.

\begin{figure*}[t]
  \centering
  
  \begin{subfigure}{0.495\textwidth}
    \centering
    \includegraphics[width=\linewidth]{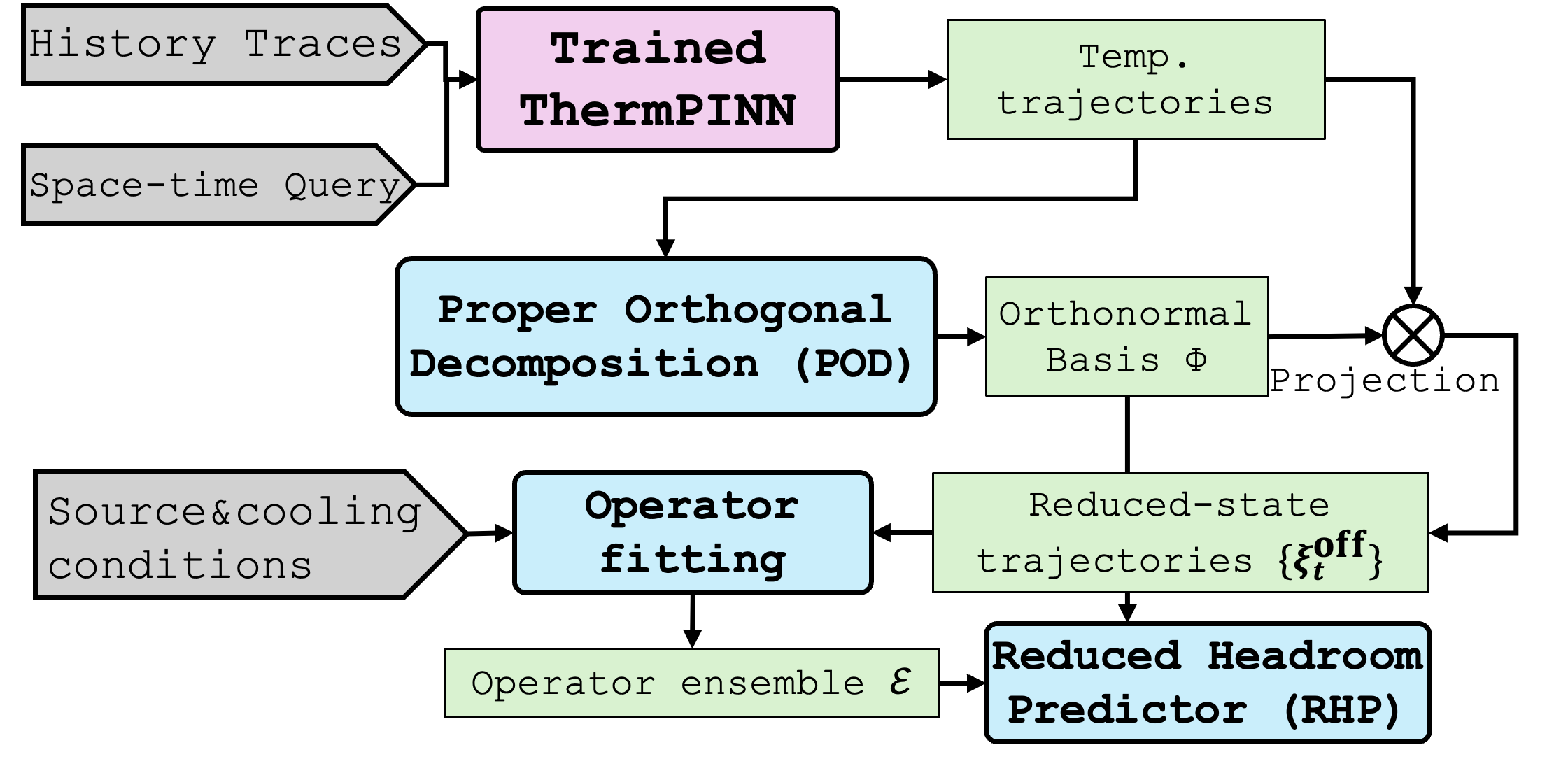}
    \caption{Offline RHP construction.}
    \label{fig:rhp-construction}
  \end{subfigure}\hfill
  \begin{subfigure}{0.495\textwidth}
    \centering
    \includegraphics[width=\linewidth]{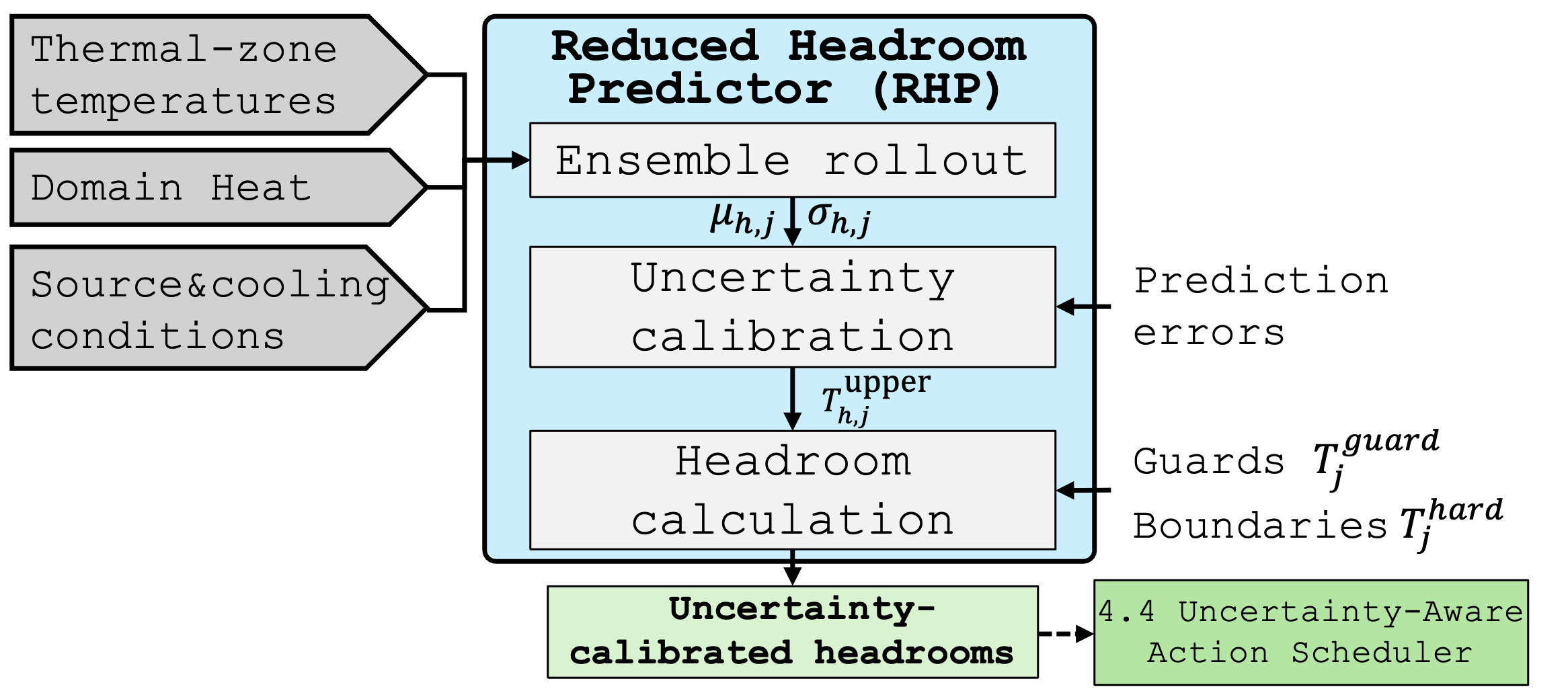}
    \caption{Online RHP prediction.}
    \label{fig:rhp-online}
  \end{subfigure}
  \caption{RHP construction and prediction. Offline calibration produces an orthonormal basis and operator ensemble. Online prediction converts candidate heat inputs and runtime observations into uncertainty-calibrated headrooms.}
  
  \label{fig:rhp-workflow}
\end{figure*}
\subsection{PDE-Constrained Headroom Predictor}
\label{sec:shared-headroom-predictor}

The scheduler needs to compare how candidate actions consume shared thermal headroom. This requires accounting for retained heat from previous work and heat propagation across domains under each action.

Theoretically, heat propagation across the shared package is governed by a \textit{transient heat-transfer PDE}~\cite{huang2006hotspot,sridhar2010_3dice}. CPU, GPU, and RAM activity introduces distributed heat sources, whose effects persist across control intervals. The unknown is the coupled temperature field $T(\mathbf{x},\tau)$ at location $\mathbf{x}$ and time $\tau$ over the package domain $\Omega$. The PDE relates its temporal evolution to spatial heat conduction and source injection
\begin{equation}
  \rho(\mathbf{x})c_p(\mathbf{x})
  \frac{\partial T(\mathbf{x},\tau)}{\partial \tau}
  =\nabla\!\cdot\!\left(k(\mathbf{x},T)\nabla T(\mathbf{x},\tau)\right)
  +q(\mathbf{x},\tau).
  \label{eq:thermal-pde}
\end{equation}
Here, $\rho$, $c_p$, $k$, and $q$ denote density, specific heat capacity, thermal conductivity, and distributed heat sources, respectively. Interface and boundary conditions enforce temperature and heat-flux continuity and ambient heat transfer. The preceding field initializes the next interval.

The predictor has two components. (i) Offline, \textit{ThermPINN} identifies coupled thermal dynamics and generates trajectories for reduced operators during platform calibration. Re-identification follows changes to the hardware or cooling configuration. (ii) Online, the \textit{Reduced Headroom Predictor (RHP)} advances thermal states with these operators and produces uncertainty-calibrated headroom estimates for scheduling, enabling repeated candidate evaluation with \textit{low computational overhead} ($\sim$11.5 ms).

\subsubsection{Offline Thermal Identification and RHP Construction}
Eq.~\ref{eq:thermal-pde} defines the physics, while sensor temperatures and domain-power and cooling histories identify effective thermal properties, spatial heat-source parameters, and internal thermal states. Sparse sensors leave internal temperatures partially observed, and imperfectly identified properties introduce prediction uncertainty~\cite{stuart2010inverse,yang2021bpinns}. ThermPINN identifies field dynamics from these observations under heat-transfer constraints, while the RHP combines ensemble prediction and residual calibration to produce thermal-zone temperature upper bounds.

ThermPINN with weights $\theta$, takes \textit{History Traces} $\mathcal{H}_{\tau}$, containing sensor temperatures, domain powers, and cooling states, and a \textit{Space-Time Query} $(\mathbf{x},\tau)$. Its output is the estimated temperature $T_{\theta}(\mathbf{x},\tau)$ at the queried location and time. Space-time queries reconstruct the field and generate trajectories for RHP construction. Its two neural components comprise a \textit{causal history encoder} that represents retained heat from the history traces and a \textit{spatiotemporal decoder} that combines this representation with the query to estimate temperature.
The input-output mapping is
\begin{equation}
\begin{aligned}
\mathbf{z}_{\mathrm{hist}}
=\operatorname{Enc}_{\theta}\!\left(\mathcal{H}_{\tau}\right),\quad
T_{\theta}(\mathbf{x},\tau)
=\operatorname{Dec}_{\theta}\!\left(\mathbf{x},\tau,\mathbf{z}_{\mathrm{hist}}\right).
\end{aligned}\nonumber
\label{eq:thermpinn-input-output}
\end{equation}
Here, $\mathcal{H}_{\tau}=\{\mathbf{y}_s,\mathbf{p}_s,v_s\}_{s\in\mathcal{W}_{\tau}}$, with $\mathcal{W}_{\tau}$ denoting the history window ending at query time $\tau$, and $\mathbf{z}_{\mathrm{hist}}$ is the encoded history representation. At each sample time $s$, $\mathbf{y}_s$ contains sensor temperatures, $\mathbf{p}_s$ contains domain powers, and $v_s$ denotes the cooling state, such as the fan operating state.

\noindent\textbf{Training of ThermPINN.}
In Fig.~\ref{fig:thermpinn-training}, training processing includes sensor measurements with heat-transfer constraints to identify the temperature field, effective thermal properties, and spatial heat-source parameters~\cite{raissi2019pinn}. Ambient temperature and idle and external heat specify the cooling and source conditions. We minimize the loss
\begin{equation}
\resizebox{0.98\columnwidth}{!}{$\displaystyle
\mathcal{L}_{\mathrm{total}}={}
\lambda_y\mathcal{L}_{\mathrm{sen}}
+\lambda_f\mathcal{L}_{\mathrm{pde}}
+\lambda_b\mathcal{L}_{\mathrm{bound}}
+\lambda_i\mathcal{L}_{\mathrm{intf}}
+\lambda_0\mathcal{L}_{\mathrm{init}}
+\lambda_r\mathcal{L}_{\mathrm{reg}}.
$}
\label{eq:design-thermpinn-loss}\nonumber
\end{equation}
Here, $\mathcal{L}_{\mathrm{sen}}$ measures temperature error at sensor locations, and $\mathcal{L}_{\mathrm{pde}}$ is the mean-squared residual of Eq.~\ref{eq:thermal-pde}, evaluated on $T_{\theta}$ using automatic differentiation. The terms $\mathcal{L}_{\mathrm{bound}}$, $\mathcal{L}_{\mathrm{intf}}$, and $\mathcal{L}_{\mathrm{init}}$ penalize violations of cooling, material continuity, and initial-state conditions, respectively, while $\mathcal{L}_{\mathrm{reg}}$ regularizes the fitted parameters. Each loss is normalized by its characteristic physical scale and balanced by its corresponding nonnegative weight.

\noindent\textbf{Generation of Reduced Headroom Predictor. } Fig.~\ref{fig:rhp-construction} summarizes the offline construction. After training ThermPINN, we generate temperature trajectories under representative source and cooling conditions to construct the RHP. Let $\mathbf{T}_{\theta,t}\in\mathbb{R}^{N}$ collect ThermPINN outputs at $N$ spatial locations at time $t\Delta t$. Proper Orthogonal Decomposition (POD) extracts an orthonormal basis $\Phi\in\mathbb{R}^{N\times r}$ containing $r\ll N$ dominant spatial modes.
The temperature fields are projected onto the reduced basis as $\boldsymbol{\xi}^{\mathrm{off}}_t=\Phi^\top\mathbf{T}_{\theta,t}$.
Here, $\boldsymbol{\xi}^{\mathrm{off}}_t$ is the reduced state obtained from the offline ThermPINN. We fit an ensemble $\mathcal{E}$ of state operators conditioned on the cooling state to these projected trajectories using distinct bootstrap weights to capture variation in the identified dynamics. The resulting basis and operators support compact state updates during online candidate evaluation.

\subsubsection{Online Headroom Prediction}
For $t$ indexing control intervals of duration $\Delta t$, $\tau$ denotes continuous time, and interval $t$ spans $[t\Delta t,(t+1)\Delta t)$. The subscript $\Delta$ in $A_\Delta$ and $B_\Delta$ indicates a one-interval update.
In Fig.~\ref{fig:rhp-online}, each interval starts from a sensor-corrected state $\boldsymbol{\xi}_t$. The ensemble $\mathcal{E}$ rolls out each candidate sequence using
\begin{equation}
\boldsymbol{\xi}_{t+1}
=A_{\Delta}(v_t)\boldsymbol{\xi}_t+B_{\Delta}(v_t)\mathbf{p}_t+\mathbf{b}_t, \qquad
\widehat{\mathbf{y}}_t=C_y\boldsymbol{\xi}_t.
\label{eq:reduced-headroom-rollout}
\end{equation}
$A_\Delta(v_t)$, $B_\Delta(v_t)$ are the state-transition and heat-input mapping matrices. The vectors $\mathbf p_t$ and $\widehat{\mathbf y}_t$ contain candidate heat inputs and predicted sensor temperatures, respectively. Here, $\mathbf{b}_t$ aggregates ambient, idle, external-heat, and residual contributions. The sensor sampling operator $S_y$ gives $C_y=S_y\Phi$. Each step advances the preceding predicted state.

At horizon $h$ and zone $j$, ensemble temperatures yield mean $\mu_{h,j}$ and standard deviation $\sigma_{h,j}$. Observed errors supply positive-bias correction $\delta^+_{h,j}$ and one-sided conformal quantile $\eta_{h,j}$. With fixed spread multiplier $\kappa$, the upper bound and headrooms are
\begin{equation}
\begin{aligned}
T^{\mathrm{upper}}_{h,j}
&=\mu_{h,j}+\delta^+_{h,j}+\kappa\sigma_{h,j}+\eta_{h,j},\\
H^{\mathrm{safe}}_{h,j}
&=T_j^{\mathrm{guard}}-T^{\mathrm{upper}}_{h,j},\,
H^{\mathrm{shared}}_h
&=\min_j H^{\mathrm{safe}}_{h,j},\\
H^{\mathrm{hard}}_h
&=\min_j\{T_j^{\mathrm{hard}}-T^{\mathrm{upper}}_{h,j}\}.
\end{aligned}
\label{eq:shared-safe-headroom}
\end{equation}
Here, $T_j^{\mathrm{guard}}$ and $T_j^{\mathrm{hard}}$ are the software guard and hard safety boundary. These headrooms support scheduler screening and planning.

\begin{figure}[t]
\centering
\begin{algorithm}[H]
\caption{Uncertainty-aware action scheduling}
\label{alg:action-scheduler}
\small
\begin{algorithmic}[1]
\State \textbf{Input} $s_t$, $a_{\mathrm{cur}}$, $m$, candidate action set $\mathcal A$, beam width $K$
\State \textbf{Output} Action for the next control interval
\State Update RHP residuals $R_m$ from observations
\State $\mathcal B\gets\{\emptyset\}$
\For{$h=1,\ldots,H_{\mathrm{plan}}$}
  \State $\mathcal Q\gets\{\boldsymbol{\pi}\mathbin{\|}a\mid\boldsymbol{\pi}\in\mathcal B,\ a\in\mathcal A\}$
  \State \textsc{Score} each candidate and discard infeasible ones
  \State If $\mathcal Q$ is empty, return fallback
  \State $\mathcal B\gets$ best $K$ by $J_m$, breaking ties by final headroom
\EndFor
\State \Return first action of the best sequence in $\mathcal B$
\Procedure{Score}{$\boldsymbol{\pi}$}
  \State Advance serving states, compile heat, and run the RHP
  \State On invalid predictions, abort scheduling and use fallback
  \State Reject out-of-range work or any $H_h^{\mathrm{hard}}<0$
  \State Hold the final action to find the first guard crossing $k$
  \State If in normal operation, $R_m\gets w_\tau k\Delta t$
  \State If recovering headroom, $R_m\gets-w_o\sum_{h=1}^{|\boldsymbol{\pi}|}[-H_h^{\mathrm{shared}}]_+$
  \State \Return $J_m$ from Eq.~\ref{eq:scheduling-objective} using predicted service metrics
\EndProcedure
\end{algorithmic}
\end{algorithm}

\end{figure}

\subsection{Uncertainty-Aware Action Scheduler}
\label{sec:controller}

The scheduler balances serving quality with thermal headroom. Let $\boldsymbol{\pi}=(a_t,\ldots,a_{t+H_{\mathrm{plan}}-1})$ denote a candidate action sequence over $H_{\mathrm{plan}}$ control intervals. For $h=1,\ldots,H_{\mathrm{plan}}$, action $a_{t+h-1}=(\mathbf{h}_{t+h-1},c_{t+h-1})$ pairs a hardware profile $\mathbf{h}_{t+h-1}$ with a scheduled-token cap $c_{t+h-1}$. Headroom at step $h$ is predicted after applying actions $a_t,\ldots,a_{t+h-1}$. The scheduler selects the highest-scoring feasible sequence $\boldsymbol{\pi}^{\star}$ using
\begin{equation}
\begin{aligned}
\boldsymbol{\pi}^{\star}
&=\arg\max_{\boldsymbol{\pi}} J_m(\boldsymbol{\pi}),\\
J_m(\boldsymbol{\pi})
&=w_g{G_{\mathrm{slo}}}-w_b{B_{\mathrm{svc}}}-w_E{E_{\mathrm{dom}}}-w_{\mathrm{sw}}{T_{\mathrm{sw}}}+R_m,\\
\text{subject to}\quad
&H^{\mathrm{hard}}_h(\boldsymbol{\pi})\geq0,
\quad h=1,\ldots,H_{\mathrm{plan}}.
\end{aligned}\nonumber
\label{eq:scheduling-objective}
\end{equation}
Here, $G_{\mathrm{slo}}$ counts SLO-compliant completed tokens, $B_{\mathrm{svc}}$ is the backlog at the end of the sequence, $E_{\mathrm{dom}}$ is domain energy, and $T_{\mathrm{sw}}$ is switching time. The weights are all normalized. The thermal term $R_m$ rewards longer runway in normal operation and penalizes the cumulative shared-headroom deficit in headroom-recovery operation. The constraint also excludes sequences that cross a hard temperature boundary.

We solve this objective with a beam search; see Algorithm~\ref{alg:action-scheduler}. The compiler estimates heat from the work admitted by each action, and the RHP predicts its effect on headroom. If the current state is unchanged from the previous interval, the scheduler reuses the previous action.
Otherwise, beam search extends candidate sequences one interval at a time and keeps the $K$ highest-scoring feasible candidates at each depth. In the algorithm, $s_t$, $a_{\mathrm{cur}}$, and $m$ denote the observed state, currently applied action, and operating regime, and $\boldsymbol{\pi}\mathbin{\|}a$ appends an action to a sequence. The scoring procedure evaluates predicted service metrics and thermal headroom. For runway, it holds the final action through $H_{\mathrm{run}}$ and sets $k$ to the first step with negative shared headroom, or $H_{\mathrm{run}}+1$ if no crossing occurs. Headroom-recovery operation starts when shared headroom becomes negative and persists until it exceeds the recovery margin. The scheduler applies only the first action of the selected sequence and updates its decision after the next observation. Fallback restores default serving under a validated conservative hardware profile.

\section{Implementation}
\label{sec:implementation}

\begin{figure}[t]
\centering

\includegraphics[width=0.8\columnwidth]{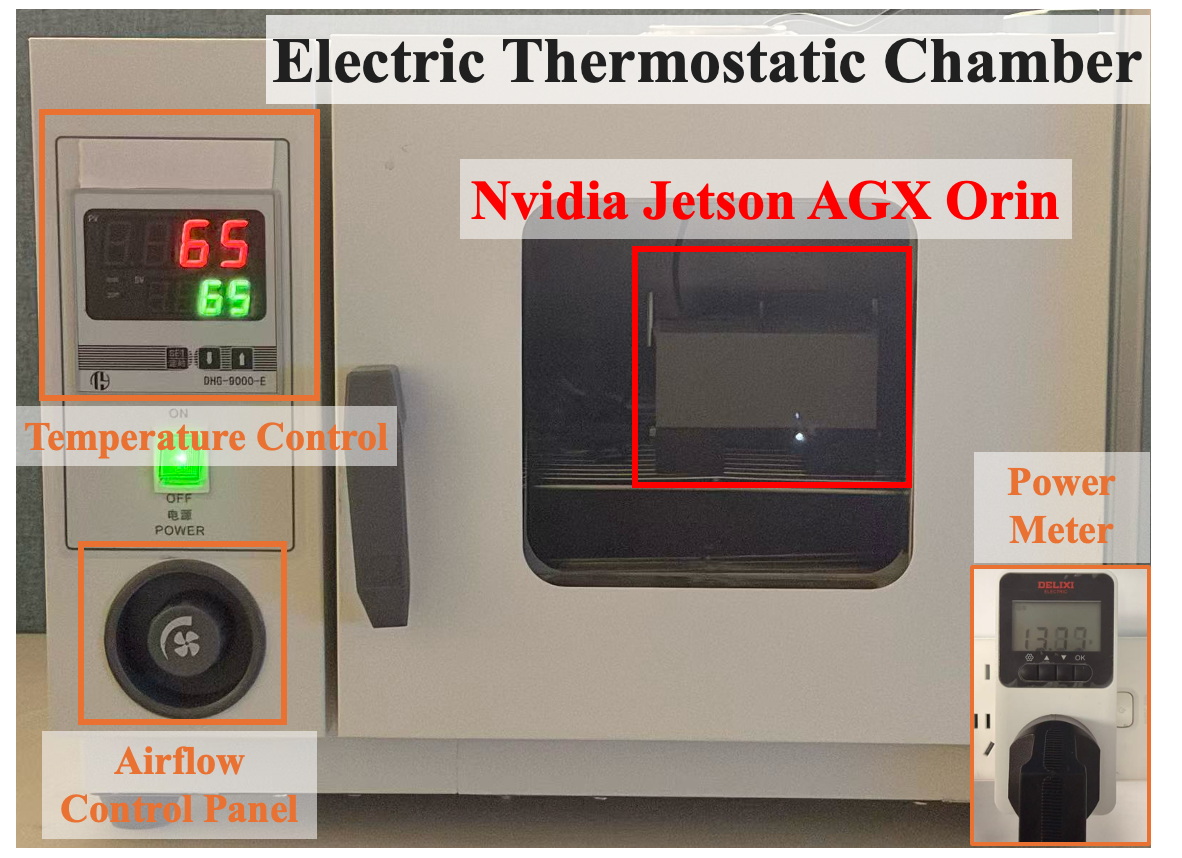}
\caption{Temperature-controlled testbed in electric thermostatic chamber for Nvidia Jetson AGX Orin experiments.}
\label{fig:evaluation-testbed}

\end{figure}

\noindent\textbf{Prototype and integration.} We implement \sysn{} as an asynchronous controller atop vLLM~\cite{kwon2023vllm}. A serving adapter exports the model descriptor, scheduled prefill and decode work, sequence lengths, batch occupancy, KV-cache use, and remaining work without blocking vLLM execution.
A platform adapter samples thermal zones, utilization, RAM activity, rail power, ambient temperature, and cooling state, represented by fan state in this prototype, every $50\,\mathrm{ms}$. Both adapters publish timestamped snapshots, and the controller accepts this synchronized pair.
The prototype hardware profiles specify CPU, GPU, and RAM frequencies. We construct candidate actions by pairing platform-supported hardware profiles with scheduled-token caps.

\noindent\textbf{ThermPINN implementation.} We train ThermPINN on 465,300 sensor temperatures, domain power, ambient temperature, and cooling state, about 267 device hours. The corpus comprises four trajectories spanning four cooling conditions and 20 workload classes on three testbeds. We split the corpus by complete trajectory into training, validation, and test sets comprising 74.7\%, 15.3\%, and 10.0\%, respectively. Sensor temperatures supervise the predicted field, while sampled interior, boundary, interface, and initial-time points enforce the PDE and associated conditions. ThermPINN can be trained with ease on an Apple MacBook in two hours.

\noindent\textbf{Online prediction and scheduling.} We construct an RHP from ThermPINN's identified thermal dynamics using POD and reduced-operator fitting.
The $1.5\,\mathrm{MB}$ RHP bundle contains reduced operators conditioned on cooling state and residuals for calibration. A dedicated worker runs every $2\,\mathrm{s}$ for a $64\,\mathrm{s}$ forecasting window to correct the RHP state and search three control intervals, using the calibrated serving-transition catalogue to predict serving behavior.
Telemetry, prediction, search, or failures trigger conservative fallback, while independent safety controls remain active and failure or exit restores the startup platform state.
% Beam-search planning takes $34.5\,\mathrm{ms}$.
The LLM-to-Heat Compiler, RHP, and scheduler took 18.13\,ms, 24.36\,ms, and 11.5\,ms on average, respectively, corresponding to 0.91\%, 1.22\%, and 0.58\% of the 2\,s control interval.

\section{Evaluation}
\label{sec:evaluation}

\begin{table}[t]
\centering

\caption{LLM workloads and serving targets}
\label{tab:evaluation-workloads}
\setlength{\tabcolsep}{2pt}
\resizebox{\columnwidth}{!}{%
\begin{tabular}{lccccccc}
\toprule
Workload & Model & Prompt & Input & Output & Interarrival & TTFT & TPOT \\
& & Trace & tokens & tokens & s/request & s & ms \\
\midrule
Stress (default) & Qwen3-14B & WikiText-103& 511 & 128 & 2.50 & 6& 230\\
Chat & Llama-3.1-8B& OASST1& 254 & 64 & 1.33 & 0.4& 160\\
Code & DeepSeek-6.7B& CodeSearchNet& 512 & 256 & 5.00 & 0.5& 300\\
Summary & Gemma-3-27B & GovReport& 1535 & 64 & 10.00 & 5& 650\\
\bottomrule
\end{tabular}
}

\end{table}

In this section, we evaluate whether \sysn{} preserves sustained LLM serving performance under limited shared-package thermal headroom.
We measure service quality and performance under thermal constraints, validate the compiler and headroom predictor separately, ablate the contribution of each component, and conduct a case study on heat accumulation inside a vehicle trunk. The source code of \sysn{} is publicly available in an anonymous repository.\footnote{\url{https://anonymous.4open.science/r/ThermE-codebase-D37B/}}

\begin{figure*}[t]
\centering

\begin{subfigure}[t]{0.235\textwidth}
\centering
\includegraphics[width=\linewidth]{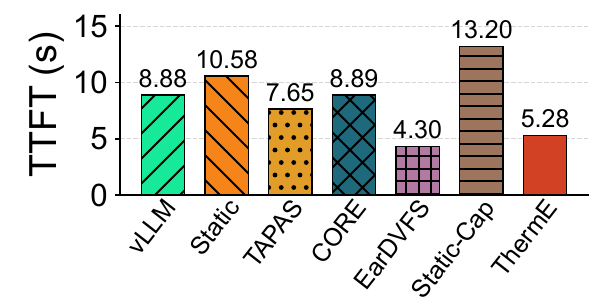}
\caption{TTFT ($\downarrow$)}
\label{fig:serving-quality-ttft}
\end{subfigure}\hfill
\begin{subfigure}[t]{0.235\textwidth}
\centering
\includegraphics[width=\linewidth]{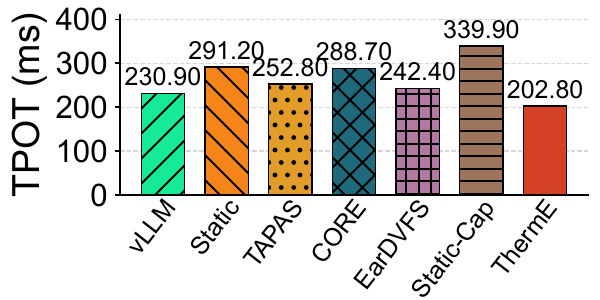}
\caption{TPOT ($\downarrow$)}
\label{fig:serving-quality-tpot}
\end{subfigure}\hfill
\begin{subfigure}[t]{0.235\textwidth}
\centering
\includegraphics[width=\linewidth]{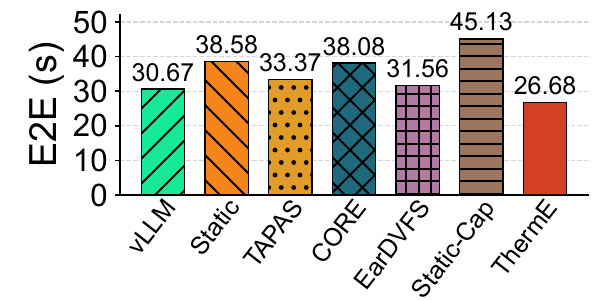}
\caption{end-to-end latency ($\downarrow$)}
\label{fig:serving-quality-e2e}
\end{subfigure}\hfill
\begin{subfigure}[t]{0.235\textwidth}
\centering
\includegraphics[width=\linewidth]{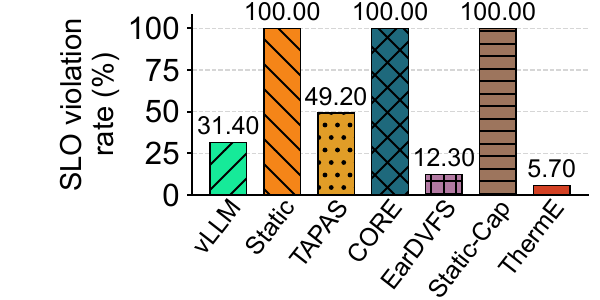}
\caption{SLO violation rate ($\downarrow$)}
\label{fig:serving-quality-slo}
\end{subfigure}
\caption{Serving quality under the default setup.}
\label{fig:serving-quality}

\end{figure*}

\subsection{Evaluation Setup}
\label{sec:evaluation-setup}

\noindent\textbf{Testbed.} We deploy \sysn{} on an Nvidia Jetson AGX Orin with 64\,GB of unified RAM and a 12-core Arm CPU, placed in a temperature-controlled electric thermostatic chamber,
see Fig.~\ref{fig:evaluation-testbed}.
The system runs on Ubuntu 22.04, with JetPack-R36.5.2 and vLLM-0.19.0. It controls temperature from 10 to 300\C with a 1\C error.

\noindent\textbf{Models and workloads.} We use four LLMs from 6.7B to 27B parameters.
The workload set combines interactive chat, code generation, long-context summarization, and fixed long-decode stress prompts.
Trace replay preserves the measured prompt lengths, output lengths, arrival order, and interarrival times. Every policy receives the same request sequence and random seed.
Table~\ref{tab:evaluation-workloads} summarizes the evaluated workloads.
Unless otherwise specified, the default setup is the Stress task with the model and request configuration, with a chamber temperature of 65\C and a cooling state of 60\% fan.

\begin{figure}[t]
\centering

\begin{subfigure}[t]{0.49\linewidth}
\centering
\includegraphics[width=\linewidth]{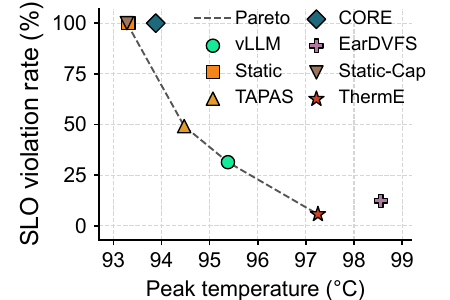}
\caption{Peak temperature versus SLO violations ($\downarrow$)}
\label{fig:thermal-pareto-slo}
\end{subfigure}\hfill
\begin{subfigure}[t]{0.49\linewidth}
\centering
\includegraphics[width=\linewidth]{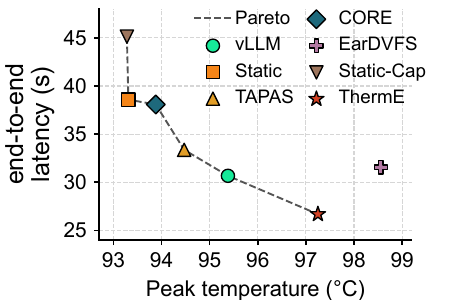}
\caption{Peak temperature versus end-to-end latency ($\downarrow$)}
\label{fig:thermal-pareto-e2e}
\end{subfigure}
\caption{Thermal--serving tradeoffs under the default setup. Dashed lines connect the non-dominated Pareto frontier when both axes are minimized.}
\label{fig:thermal-serving-pareto}

\end{figure}

\noindent\textbf{Baselines.}
We compare \sysn{} against representative LLM serving and controlling baselines:
\begin{itemize}[leftmargin=*,itemsep=0pt,topsep=2pt]
\item \textbf{vLLM}~\cite{kwon2023vllm} uses the platform's default CPU, GPU, and RAM governors together with vendor thermal throttling.
\item \textbf{Static} fixes the validation-selected hardware profile, specifying joint CPU, GPU, and RAM frequencies greedily that meet the TTFT and TPOT targets, without runtime adaptation.
\item \textbf{TAPAS}~\cite{stojkovic2025tapas} is for thermal-/power-aware LLM inference in clouds. We adapt only its linear regression method to predict temperature from ambient temperature and device load. We select among the same hardware profiles based on predicted thermal and power headroom.
\item \textbf{CORE}~\cite{zhang2026core} profiles joint CPU, GPU, and RAM frequency combinations offline and applies a precomputed combination for each prefill or decode phase under the same latency targets.
\item \textbf{EarDVFS}~\cite{kwak2025eardvfs} uses environment-aware reinforcement learning to set maximum-frequency boundaries from chamber ambient, package temperature, utilization, power, and frequency telemetry, while retaining the vendor governor for final clock selection.
\item \textbf{Static-Cap} applies one fixed platform power cap selected on the validation trace and retains the default governors.
\end{itemize}

\noindent\textbf{Metrics.}
We evaluate three aspects. i) Serving quality is measured by the 95th percentile (p95) of time to first token (TTFT), time per output token (TPOT), and end-to-end latency across completed requests, together with the SLO violation rate. ii) Thermal behavior is measured by peak package temperature and time to first throttle. iii) Power is averaged over the measurement interval onboard. Total device power is measured at the device input using an external power meter.
For the compiler and headroom predictor, we report domain-power error, temperature mean absolute error (MAE) and root mean squared error (RMSE).

\begin{figure}[t]
\centering

\begin{minipage}[t]{0.485\columnwidth}
\centering
\includegraphics[width=\linewidth]{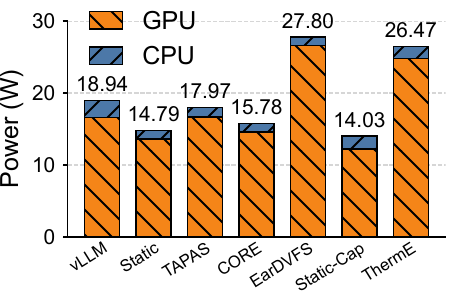}
\caption{CPU/GPU power under the default setup.}
\label{fig:average-power-breakdown}
\end{minipage}\hfill
\begin{minipage}[t]{0.485\columnwidth}
\centering
\includegraphics[width=\linewidth]{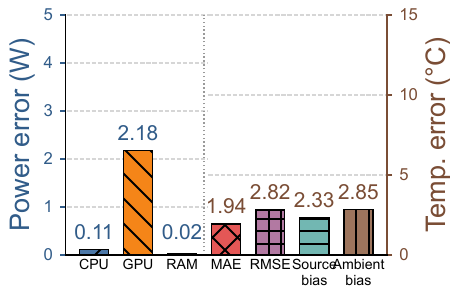}
\caption{Power \& thermal prediction error ($\downarrow$).}
\label{fig:evaluation-prediction-error}
\end{minipage}

\end{figure}

\subsection{Overall Performance}
\label{sec:evaluation-overall}

We evaluate serving quality, thermal behavior, power, and runtime overhead under the default setup. Every policy receives the same request trace and SLO targets.

\noindent\textbf{Improvement of Serving Quality.} Fig.~\ref{fig:serving-quality} compares serving quality. During the thermally steady interval, \sysn{} maintains TTFT and TPOT at 5.28\,s and 202.80\,ms.
\sysn{} reduces TTFT by 40.55\%, TPOT by 12.17\%, and end-to-end latency by 13.01\% relative to vLLM. Its SLO violation rate is 5.70\%, compared with 12.30\% for the strongest baseline.
These gains are consistent with joint selection of hardware profiles and scheduled-token caps using serving demand and predicted headroom, which allows \sysn{} to account for the service cost of each thermal-control decision.

\noindent\textbf{Thermal--serving tradeoff.} Fig.~\ref{fig:thermal-serving-pareto} relates peak temperature to SLO violations and end-to-end latency. \sysn{} first throttled after 29.70\,s, while vLLM first throttled after 63.10\,s. At this operating point, \sysn{} achieved a 5.70\% SLO violation rate and 26.68\,s end-to-end latency, compared with 31.40\% and 30.67\,s for vLLM. \sysn{} therefore lies on the non-dominated frontier in both comparisons, achieving better serving quality while accepting higher thermal exposure. This operating point reflects its scheduling objective, which balances serving progress with future headroom and screens candidate actions against hard temperature boundaries.

\noindent\textbf{Power Characterization.} Fig.~\ref{fig:average-power-breakdown} compares CPU and GPU power across baselines. \sysn{} records CPU and GPU power of 1.70\,W and 24.77\,W. Relative to vLLM, these values change by -28.42\% and +49.51\%, respectively. Baseline CPU power ranges from 1.20\,W to 2.37\,W and GPU power ranges from 12.24\,W to 26.56\,W. Together with the latency reductions, this power distribution indicates that \sysn{} sustains more GPU activity while using less CPU power than vLLM, and that we trade off CPU and GPU settings together with their effects on serving quality and shared headroom.

\label{sec:evaluation-prediction}
\noindent\textbf{Prediction Accuracy.}
We evaluate the predictions used to estimate available thermal headroom on combinations of models, requests, and chamber conditions. Fig.~\ref{fig:evaluation-prediction-error} reports CPU, GPU, and RAM power errors of 0.11\,W, 2.18\,W, and 0.02\,W, respectively, characterizing the compiler inputs used to forecast headroom consumption. The predictor achieved a temperature MAE of 1.94\C and RMSE of 2.82\C across thermal zones. For a fixed thermal threshold, temperature-prediction errors translate into per-zone headroom errors of equal magnitude. These forecasts give the scheduler a quantitative basis for comparing candidate actions before execution, while residual calibration adjusts the exposed headroom to account for prediction uncertainty.

\begin{figure}[t]
\centering

\begin{subfigure}[t]{0.48\linewidth}
\centering
\includegraphics[width=\linewidth]{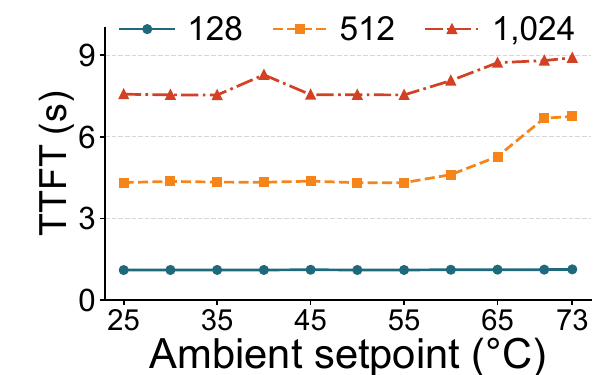}
\caption{TTFT ($\downarrow$)}
\label{fig:evaluation-ambient-ttft}
\end{subfigure}\hfill
\begin{subfigure}[t]{0.48\linewidth}
\centering
\includegraphics[width=\linewidth]{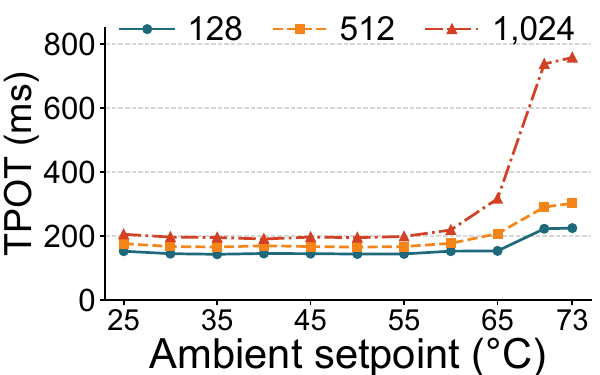}
\caption{TPOT ($\downarrow$)}
\label{fig:evaluasion-ambient-tpot}
\end{subfigure}
\caption{Ambient-temperature sensitivity with different prompt lengths.}
\label{fig:evaluation-ambient}
\end{figure}

\begin{table}[t]
\centering

\caption{Comparision of \sysn{} to the lowest and highest baselines for each metric on different workloads.}
\label{tab:workload-relative-performance}
\setlength{\tabcolsep}{3pt}
\resizebox{\columnwidth}{!}{%
\begin{tabular}{lcccc}
\toprule
Workload & $\Delta$TTFT & $\Delta$TPOT & $\Delta$Peak temp. & $\Delta$GPU power \\
& (\%) & (\%) & (\C) & (\%) \\
\midrule
Chat & $-33.33 / -95.13$ & $-28.56 / -42.34$ & $+1.00 / -0.72$ & $+126.37 / +66.58$ \\
Code & $-9.62 / -96.25$ & $-5.46 / -30.11$ & $+1.28 / +0.15$ & $+58.14 / +9.72$ \\
Summary & $+11.06 / -88.95$ & $-41.33 / -85.66$ & $+4.06 / -1.06$ & $+94.05 / -1.78$ \\
\bottomrule
\end{tabular}%
}

\end{table}

\subsection{Sensitivity Analysis}
\label{sec:evaluation-sensitivity}

We examine sensitivity to ambient temperature, prompt length, cooling state, and dataset. Each experiment changes the corresponding setting while retaining the other default settings.

\noindent\textbf{Impact of Ambient Temperature Setpoint.} We evaluated \sysn{} with Qwen3-14B-AWQ at ambient setpoints from 30\C to 73\C with 100\% fan speed. Fig.~\ref{fig:evaluation-ambient} shows that increasing ambient temperature raised serving latency, with larger increases above 60\C. At a 512-token prompt length, TTFT increased from 4.36\,s at 30\C to 6.69\,s at 70\C, while TPOT increased from 167.36 to 290.44\,ms.
The other prompt lengths exhibited a similar temperature-dependent increase.
Thus, \sysn{} maintained broadly stable serving at lower setpoints and continued processing requests as higher ambient temperatures reduced capacity. Its ambient-conditioned headroom forecasts expose the shrinking thermal margin to the scheduler, allowing action selection to account for environmental heating.

\begin{table}[t]
\centering

\caption{Percentage changes relative to the complete system.}
\label{tab:evaluation-ablation}
\setlength{\tabcolsep}{3pt}
\resizebox{0.9\columnwidth}{!}{%
\begin{tabular}{lccc}
\toprule
Variant & TTFT(\%) & TPOT(\%) & SLO vio. rate (\%) \\
\midrule
Without Compiler & +11.02 & +14.32 & +13.57 \\
Without Coupling & +14.01 & +12.57 & +14.23 \\
Without Uncertainty & +11.05 & +11.63 & +13.02 \\
Greedy & +13.18 & +13.19 & +15.63 \\
Full numerical solver & +8.70 & +26.68 & +48.20 \\
\bottomrule
\end{tabular}%
}

\end{table}

\noindent\textbf{Impact of Prompt Length.} We also compared 128-, 512-, and 1,024-token prompts in Fig.~\ref{fig:evaluation-ambient}.
At 30\C, it shows that increasing prompt length from 128 to 1,024 tokens raised TTFT from 1.10 to 7.55\,s and TPOT from 144.76 to 196.67 ms. Longer prompts require more prefill work and enlarge the context accessed during decoding, increasing the work in both phases. The same trend held at higher setpoints, with a wider latency gap for longer prompts. These results show that \sysn{} supported all three input lengths with different serving costs. Its compiler represents prompt-dependent compute and memory work, enabling the scheduler to account for the corresponding heat demand when selecting hardware profiles and scheduled-token caps.

\noindent\textbf{Performance across workloads.} We then test the \sysn{} on Chat, Code, and Summary tasks in Table~\ref{tab:evaluation-workloads}, while other settings remain at their default values. Table~\ref{tab:workload-relative-performance} compares TTFT, TPOT, peak temperature, and GPU power with the best and worst baselines for each metric. \sysn{} reduced TPOT relative to the strongest baseline by 28.56\%, 5.46\%, and 41.33\% for Chat, Code, and Summary, respectively, showing a consistent decoding benefit across workloads. For Summary, \sysn{} reduced TTFT by up to 88.95\% across baselines, although it remained 11.06\% higher than the best baseline, suggesting a tradeoff in which headroom-aware scheduling sacrifices some prefill speed to sustain faster decoding. \sysn{} uses model and request descriptors to balance serving progress against shared-headroom consumption.

\subsection{Ablation Studies}
\label{sec:evaluation-ablation}

We evaluate five variants under the same action space and thermal guard. \textit{Without Compiler} replaces the package activity signature with GPU utilization and total power. \textit{Without Coupling} uses independent per-zone temperature predictors. \textit{Without Uncertainty} plans from the ensemble mean without a conformal margin. \textit{Greedy} replaces horizon search with a one-step decision. \textit{Full numerical solver} replaces the reduced thermal predictor with a full numerical solver.

Table~\ref{tab:evaluation-ablation} shows that removing the compiler, thermal coupling, or uncertainty margin increased TTFT, TPOT, and SLO violations by 11.02--14.32\%. These components jointly account for workload-dependent heat generation, cross-domain propagation, and prediction uncertainty, supporting better serving decisions. Greedy increased TTFT and TPOT by 13.18\% and 13.19\%, respectively, and SLO violations by 15.63\%, demonstrating the benefit of anticipating how current actions consume future headroom. The full numerical solver increased TTFT by 8.70\%, TPOT by 26.68\%, and SLO violations by 48.20\%, yielding the largest decoding and SLO penalties. The reduced predictor's compact state updates offer a plausible explanation for its serving advantage by limiting candidate-evaluation computation. Together, these results support combining predictive headroom estimation with multi-interval planning for sustained serving under thermal constraints.

\subsection{Sustained Vehicle-Trunk Thermal Case Study}
\label{sec:evaluation-vehicle-case-study}

\begin{figure}[t]
\centering
\includegraphics[width=\linewidth]{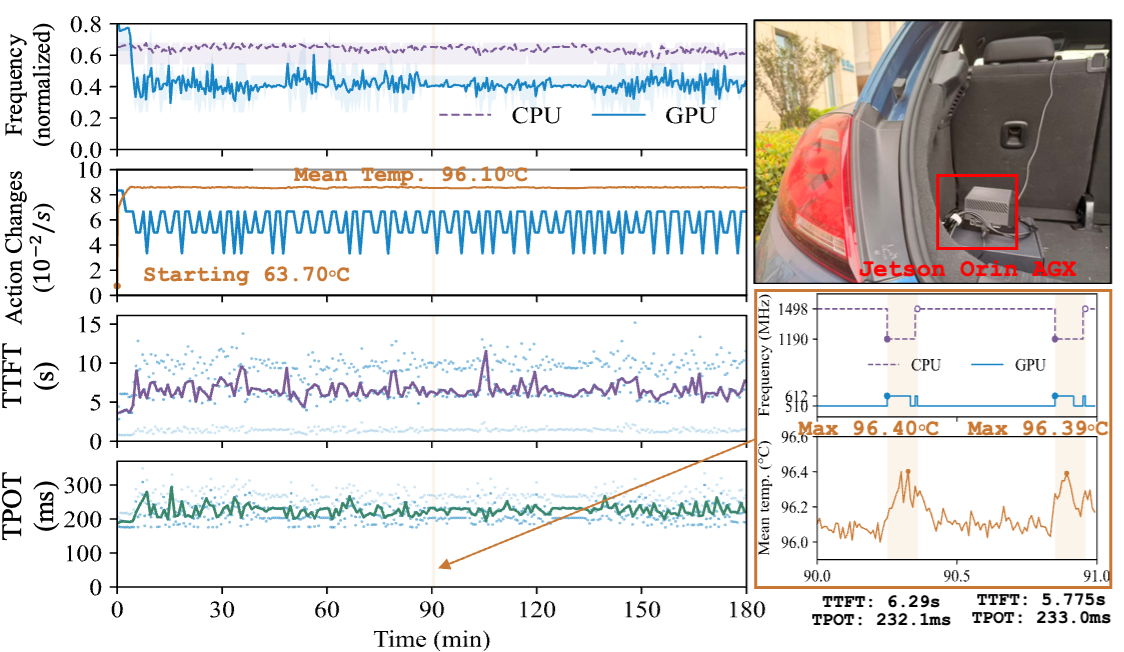}
\caption{Case study results of 180-minute sustained operations, including CPU/GPU frequencies, action-change rate with temperature, TTFT, and TPOT. The bottom-right details frequency transitions and temperature during 90--91 min.}
\label{fig:vehicle-therme}

\end{figure}

We deploy \sysn{} under sustained heat accumulation in vehicle-trunk operation. Fig.~\ref{fig:vehicle-therme} presents a 180-minute run of Qwen3-14B-AWQ with 512-token input prompts and 128-token outputs, with eight concurrent requests. The device fan duty was 68\%, and temperatures and frequencies were sampled every 0.5 s.

Fig.~\ref{fig:vehicle-therme} shows sustained serving with rising temperature, and finally the SoC is near 96.1\C. The CPU frequency stayed around 0.65, while the GPU frequency was near 0.4.
Across the run, \sysn{} completed 2,448 requests and delivered 29.01 output tokens/s, with TTFT of 11.10\,s and TPOT of 273.98 ms. Latency varied throughout the run but did not increase.
The action-change trace shows continued adaptation throughout. The controller applied approximately 3.41 changes per minute. The 90--91-minute inset shows two recurring transitions in which CPU frequency decreased from approximately 1,498 to 1,190 MHz while GPU frequency increased from 510 to 612 MHz. Both frequencies subsequently returned to their preceding levels. The corresponding temperature excursions peaked at 96.40\C and 96.39\C before declining toward 96.1\C. Together, these traces show sustained request processing with recurring CPU/GPU adjustments under persistent thermal pressure.

\section{Related Work}
\label{sec:related}

\noindent\textbf{Efficient LLM inference on heterogeneous platforms.}
vLLM improves throughput with PagedAttention~\cite{kwon2023vllm}, and POD-Attention overlaps prefill and decode attention~\cite{kamath2025podattention}. Helix optimizes placement and routing across heterogeneous GPUs~\cite{mei2025helix}. Mobile systems exploit NPUs for prefill~\cite{xu2025fastnpu}, combine GPU and NPU execution~\cite{moon2025hybe}, and coordinate CPU, GPU, and memory frequencies~\cite{zhang2026core}. \sysn{} complements these optimizations with predictive thermal control for sustained serving under limited shared headroom.

\noindent\textbf{Thermal Modeling and Headroom Prediction.}
3D-ICE models nonlinear thermal transients and heat sinks~\cite{terraneo2022_3dice}, while DeepOHeat maps power and boundary conditions to temperature fields~\cite{liu2023deepoheat}. SAU-FNO combines attention and neural operators for thermal prediction~\cite{huang2025saufno}. ASRR-PINN uses adaptive sampling for physics-informed thermal analysis~\cite{zhou2025asrrpinn}. TAPAS estimates from thermal and power histories~\cite{stojkovic2025tapas}, while ReThermal accelerates wafer-scale thermal simulation with neural networks~\cite{li2026rethermal}. \sysn{} complements these models by translating uncertainty-calibrated forecasts into shared headroom for low-overhead serving decisions.

\noindent\textbf{Thermal-Aware Resource Management.}
EarDVFS adapts mobile frequency control using thermal and cooling state~\cite{kwak2025eardvfs}. TAPAS coordinates placement, routing, and reconfiguration under cloud thermal and power constraints~\cite{stojkovic2025tapas}. ReThermal combines thermal modeling and scheduling for liquid-cooled wafer-scale LLM training~\cite{li2026rethermal}. \sysn{} extends workload-aware thermal management to shared-cooling edge SoCs, jointly controlling hardware profiles and scheduled-token caps to preserve serving quality.

\section{Conclusion}
\label{sec:conclusion}

We presented ThermE, a runtime system that manages shared thermal headroom for sustained LLM inference on edge SoCs. ThermE combines LLM-to-heat compilation, physics-constrained headroom prediction, and uncertainty-aware scheduling to coordinate hardware profiles and scheduled-token caps. Evaluation demonstrated improved serving quality under thermal constraints. This work contributes a predictive thermal resource management approach to the systems community, providing a foundation for sustained onboard intelligence in satellites, smart AI mobile devices, and other compact platforms like System-on-package (SoP) where limited cooling constrains continuous inference.

  \bibliographystyle{ACM-Reference-Format}
  \bibliography{ref}


\begin{thebibliography}{32}

%%% the Web address.

\ifx \showCODEN    \undefined \def \showCODEN     #1{\unskip}     \fi
\ifx \showISBNx    \undefined \def \showISBNx     #1{\unskip}     \fi
\ifx \showISBNxiii \undefined \def \showISBNxiii  #1{\unskip}     \fi
\ifx \showISSN     \undefined \def \showISSN      #1{\unskip}     \fi
\ifx \showLCCN     \undefined \def \showLCCN      #1{\unskip}     \fi
\ifx \shownote     \undefined \def \shownote      #1{#1}          \fi
\ifx \showarticletitle \undefined \def \showarticletitle #1{#1}   \fi
\ifx \showURL      \undefined \def \showURL       {\relax}        \fi
% invisible to TeX
\providecommand\bibfield[2]{#2}
\providecommand\bibinfo[2]{#2}
\providecommand\natexlab[1]{#1}
\providecommand\showeprint[2][]{arXiv:#2}

\bibitem[Cao et~al\mbox{.}(2025)]%
        {cao2025amali}
\bibfield{author}{\bibinfo{person}{Shiheng Cao}, \bibinfo{person}{Junmin Wu},
  \bibinfo{person}{Junshi Chen}, \bibinfo{person}{Hong An}, {and}
  \bibinfo{person}{Zhibin Yu}.} \bibinfo{year}{2025}\natexlab{}.
\newblock \showarticletitle{{AMALI}: An Analytical Model for Accurately
  Modeling {LLM} Inference on Modern {GPUs}}. In
  \bibinfo{booktitle}{\emph{Proceedings of the 52nd Annual International
  Symposium on Computer Architecture}} \emph{(\bibinfo{series}{ISCA '25})}.
  \bibinfo{publisher}{ACM}, \bibinfo{address}{New York, NY, USA},
  \bibinfo{pages}{1495--1508}.
\newblock


\bibitem[Feng et~al\mbox{.}(2025)]%
        {feng2025windserve}
\bibfield{author}{\bibinfo{person}{Jingqi Feng}, \bibinfo{person}{Yukai Huang},
  \bibinfo{person}{Rui Zhang}, \bibinfo{person}{Sicheng Liang},
  \bibinfo{person}{Ming Yan}, {and} \bibinfo{person}{Jie Wu}.}
  \bibinfo{year}{2025}\natexlab{}.
\newblock \showarticletitle{WindServe: Efficient Phase-Disaggregated {LLM}
  Serving with Stream-based Dynamic Scheduling}. In
  \bibinfo{booktitle}{\emph{Proceedings of the 52nd Annual International
  Symposium on Computer Architecture}} \emph{(\bibinfo{series}{ISCA '25})}.
  \bibinfo{publisher}{ACM}, \bibinfo{address}{New York, NY, USA},
  \bibinfo{pages}{1283--1295}.
\newblock


\bibitem[Horak et~al\mbox{.}(2017)]%
        {horak2017cabin}
\bibfield{author}{\bibinfo{person}{Johannes Horak}, \bibinfo{person}{Ivo
  Schmerold}, \bibinfo{person}{Kurt Wimmer}, {and} \bibinfo{person}{G{\"u}nther
  Schauberger}.} \bibinfo{year}{2017}\natexlab{}.
\newblock \showarticletitle{Cabin air temperature of parked vehicles in summer
  conditions: life-threatening environment for children and pets calculated by
  a dynamic model}.
\newblock \bibinfo{journal}{\emph{Theoretical and Applied Climatology}}
  \bibinfo{volume}{130} (\bibinfo{year}{2017}), \bibinfo{pages}{107--118}.
\newblock


\bibitem[Huang et~al\mbox{.}(2006)]%
        {huang2006hotspot}
\bibfield{author}{\bibinfo{person}{Wei Huang}, \bibinfo{person}{Shougata
  Ghosh}, \bibinfo{person}{Sivakumar Velusamy}, \bibinfo{person}{Karthik
  Sankaranarayanan}, \bibinfo{person}{Kevin Skadron}, {and}
  \bibinfo{person}{Mircea~R. Stan}.} \bibinfo{year}{2006}\natexlab{}.
\newblock \showarticletitle{{HotSpot}: A Compact Thermal Modeling Methodology
  for Early-Stage {VLSI} Design}.
\newblock \bibinfo{journal}{\emph{IEEE Transactions on Very Large Scale
  Integration (VLSI) Systems}} \bibinfo{volume}{14}, \bibinfo{number}{5}
  (\bibinfo{year}{2006}), \bibinfo{pages}{501--513}.
\newblock


\bibitem[Huang et~al\mbox{.}(2025)]%
        {huang2025saufno}
\bibfield{author}{\bibinfo{person}{Zhen Huang}, \bibinfo{person}{Hong Wang},
  \bibinfo{person}{Wenkai Yang}, \bibinfo{person}{Muxi Tang},
  \bibinfo{person}{Depeng Xie}, \bibinfo{person}{Ting-Jung Lin},
  \bibinfo{person}{Yu Zhang}, \bibinfo{person}{Wei~W. Xing}, {and}
  \bibinfo{person}{Lei He}.} \bibinfo{year}{2025}\natexlab{}.
\newblock \showarticletitle{Self-Attention to Operator Learning-based {3D-IC}
  Thermal Simulation}. In \bibinfo{booktitle}{\emph{2025 62nd ACM/IEEE Design
  Automation Conference (DAC)}}. \bibinfo{publisher}{IEEE},
  \bibinfo{address}{Piscataway, NJ, USA}, \bibinfo{pages}{1--7}.
\newblock


\bibitem[{Intergovernmental Panel on Climate Change}(2021)]%
        {ipcc-ar6-wg1-spm}
\bibfield{author}{\bibinfo{person}{{Intergovernmental Panel on Climate
  Change}}.} \bibinfo{year}{2021}\natexlab{}.
\newblock \bibinfo{booktitle}{\emph{{Climate Change 2021: The Physical Science
  Basis. Summary for Policymakers}}}.
\newblock \bibinfo{type}{{T}echnical {R}eport}.
  \bibinfo{institution}{Intergovernmental Panel on Climate Change}.
\newblock
\urldef\tempurl%
\url{https://www.ipcc.ch/report/ar6/wg1/}
\showURL{%
\tempurl}


\bibitem[Jung et~al\mbox{.}(2025)]%
        {jung2025aria}
\bibfield{author}{\bibinfo{person}{Chanyoung Jung}, \bibinfo{person}{Jeho Lee},
  \bibinfo{person}{Gunjoong Kim}, \bibinfo{person}{Jiwon Kim},
  \bibinfo{person}{Seonghoon Park}, {and} \bibinfo{person}{Hojung Cha}.}
  \bibinfo{year}{2025}\natexlab{}.
\newblock \showarticletitle{{ARIA}: Optimizing Vision Foundation Model
  Inference on Heterogeneous Mobile Processors for Augmented Reality}. In
  \bibinfo{booktitle}{\emph{Proceedings of the 23rd Annual International
  Conference on Mobile Systems, Applications, and Services}}
  \emph{(\bibinfo{series}{MobiSys '25})}. \bibinfo{publisher}{ACM},
  \bibinfo{address}{New York, NY, USA}, \bibinfo{pages}{249--262}.
\newblock


\bibitem[Kamath et~al\mbox{.}(2025)]%
        {kamath2025podattention}
\bibfield{author}{\bibinfo{person}{Aditya~K. Kamath}, \bibinfo{person}{Ramya
  Prabhu}, \bibinfo{person}{Jayashree Mohan}, \bibinfo{person}{Simon Peter},
  \bibinfo{person}{Ramachandran Ramjee}, {and} \bibinfo{person}{Ashish
  Panwar}.} \bibinfo{year}{2025}\natexlab{}.
\newblock \showarticletitle{{POD-Attention}: Unlocking Full Prefill-Decode
  Overlap for Faster {LLM} Inference}. In \bibinfo{booktitle}{\emph{Proceedings
  of the 30th ACM International Conference on Architectural Support for
  Programming Languages and Operating Systems, Volume 2}}
  \emph{(\bibinfo{series}{ASPLOS '25})}. \bibinfo{publisher}{ACM},
  \bibinfo{address}{New York, NY, USA}, \bibinfo{pages}{897--912}.
\newblock


\bibitem[Kwak et~al\mbox{.}(2025)]%
        {kwak2025eardvfs}
\bibfield{author}{\bibinfo{person}{Jaeheon Kwak}, \bibinfo{person}{Sangeun Oh},
  \bibinfo{person}{Jinkyu Lee}, {and} \bibinfo{person}{Insik Shin}.}
  \bibinfo{year}{2025}\natexlab{}.
\newblock \showarticletitle{{EarDVFS}: Environment-Adaptable {RL}-Based {DVFS}
  for Mobile Devices}. In \bibinfo{booktitle}{\emph{Proceedings of the 44th
  IEEE/ACM International Conference on Computer-Aided Design}}
  \emph{(\bibinfo{series}{ICCAD '25})}. \bibinfo{publisher}{IEEE},
  \bibinfo{pages}{1--9}.
\newblock


\bibitem[Kwon et~al\mbox{.}(2023)]%
        {kwon2023vllm}
\bibfield{author}{\bibinfo{person}{Woosuk Kwon}, \bibinfo{person}{Zhuohan Li},
  \bibinfo{person}{Siyuan Zhuang}, \bibinfo{person}{Ying Sheng},
  \bibinfo{person}{Lianmin Zheng}, \bibinfo{person}{Cody~Hao Yu},
  \bibinfo{person}{Joseph~E. Gonzalez}, \bibinfo{person}{Hao Zhang}, {and}
  \bibinfo{person}{Ion Stoica}.} \bibinfo{year}{2023}\natexlab{}.
\newblock \showarticletitle{Efficient Memory Management for Large Language
  Model Serving with {PagedAttention}}. In
  \bibinfo{booktitle}{\emph{Proceedings of the 29th Symposium on Operating
  Systems Principles}} \emph{(\bibinfo{series}{SOSP '23})}.
  \bibinfo{publisher}{ACM}, \bibinfo{address}{New York, NY, USA},
  \bibinfo{pages}{611--626}.
\newblock


\bibitem[Lee et~al\mbox{.}(2025)]%
        {lee2025aif}
\bibfield{author}{\bibinfo{person}{Jaeyong Lee}, \bibinfo{person}{Hyeunjoo
  Kim}, \bibinfo{person}{Sanghun Oh}, \bibinfo{person}{Myoungjun Chun},
  \bibinfo{person}{Myungsuk Kim}, {and} \bibinfo{person}{Jihong Kim}.}
  \bibinfo{year}{2025}\natexlab{}.
\newblock \showarticletitle{{AiF}: Accelerating On-Device {LLM} Inference Using
  In-Flash Processing}. In \bibinfo{booktitle}{\emph{Proceedings of the 52nd
  Annual International Symposium on Computer Architecture}}
  \emph{(\bibinfo{series}{ISCA '25})}. \bibinfo{publisher}{ACM},
  \bibinfo{address}{New York, NY, USA}, \bibinfo{pages}{529--543}.
\newblock


\bibitem[Li et~al\mbox{.}(2026)]%
        {li2026rethermal}
\bibfield{author}{\bibinfo{person}{Chengran Li}, \bibinfo{person}{Huizheng
  Wang}, \bibinfo{person}{Jiaxin Liu}, \bibinfo{person}{Jingyao Liu},
  \bibinfo{person}{Zhiheng Yue}, \bibinfo{person}{Xia Li},
  \bibinfo{person}{Shenfei Jiang}, \bibinfo{person}{Jinyi Deng},
  \bibinfo{person}{Yang Hu}, {and} \bibinfo{person}{Shouyi Yin}.}
  \bibinfo{year}{2026}\natexlab{}.
\newblock \showarticletitle{{ReThermal}: Co-Design of Thermal-Aware Static and
  Dynamic Scheduling for {LLM} Training on Liquid-Cooled Wafer-Scale Chips}. In
  \bibinfo{booktitle}{\emph{Proceedings of the 32nd IEEE International
  Symposium on High-Performance Computer Architecture}}
  \emph{(\bibinfo{series}{HPCA '26})}. \bibinfo{publisher}{IEEE},
  \bibinfo{address}{Sydney, Australia}, \bibinfo{pages}{1--15}.
\newblock


\bibitem[Li et~al\mbox{.}(2025)]%
        {li2025h2llm}
\bibfield{author}{\bibinfo{person}{Cong Li}, \bibinfo{person}{Yihan Yin},
  \bibinfo{person}{Xintong Wu}, \bibinfo{person}{Jingchen Zhu},
  \bibinfo{person}{Zhutianya Gao}, \bibinfo{person}{Dimin Niu},
  \bibinfo{person}{Qiang Wu}, \bibinfo{person}{Xin Si}, \bibinfo{person}{Yuan
  Xie}, \bibinfo{person}{Chen Zhang}, {and} \bibinfo{person}{Guangyu Sun}.}
  \bibinfo{year}{2025}\natexlab{}.
\newblock \showarticletitle{{H2-LLM}: Hardware-Dataflow Co-Exploration for
  Heterogeneous Hybrid-Bonding-Based Low-Batch {LLM} Inference}. In
  \bibinfo{booktitle}{\emph{Proceedings of the 52nd Annual International
  Symposium on Computer Architecture}} \emph{(\bibinfo{series}{ISCA '25})}.
  \bibinfo{publisher}{ACM}, \bibinfo{address}{New York, NY, USA},
  \bibinfo{pages}{194--210}.
\newblock


\bibitem[Liu et~al\mbox{.}(2023)]%
        {liu2023deepoheat}
\bibfield{author}{\bibinfo{person}{Ziyue Liu}, \bibinfo{person}{Yixing Li},
  \bibinfo{person}{Jing Hu}, \bibinfo{person}{Xinling Yu},
  \bibinfo{person}{Shinyu Shiau}, \bibinfo{person}{Xin Ai},
  \bibinfo{person}{Zhiyu Zeng}, {and} \bibinfo{person}{Zheng Zhang}.}
  \bibinfo{year}{2023}\natexlab{}.
\newblock \showarticletitle{{DeepOHeat}: Operator Learning-based Ultra-fast
  Thermal Simulation in {3D-IC} Design}. In
  \bibinfo{booktitle}{\emph{Proceedings of the 60th ACM/IEEE Design Automation
  Conference}}. \bibinfo{publisher}{IEEE}, \bibinfo{pages}{1--6}.
\newblock


\bibitem[Mei et~al\mbox{.}(2025)]%
        {mei2025helix}
\bibfield{author}{\bibinfo{person}{Yixuan Mei}, \bibinfo{person}{Yonghao
  Zhuang}, \bibinfo{person}{Xupeng Miao}, \bibinfo{person}{Juncheng Yang},
  \bibinfo{person}{Zhihao Jia}, {and} \bibinfo{person}{Rashmi Vinayak}.}
  \bibinfo{year}{2025}\natexlab{}.
\newblock \showarticletitle{Helix: Serving Large Language Models over
  Heterogeneous {GPUs} and Network via Max-Flow}. In
  \bibinfo{booktitle}{\emph{Proceedings of the 30th ACM International
  Conference on Architectural Support for Programming Languages and Operating
  Systems, Volume 1}} \emph{(\bibinfo{series}{ASPLOS '25})}.
  \bibinfo{publisher}{ACM}, \bibinfo{address}{New York, NY, USA},
  \bibinfo{pages}{586--602}.
\newblock


\bibitem[Moon et~al\mbox{.}(2025)]%
        {moon2025hybe}
\bibfield{author}{\bibinfo{person}{Seungjae Moon}, \bibinfo{person}{Junseo
  Cha}, \bibinfo{person}{Hyunjun Park}, {and} \bibinfo{person}{Joo-Young Kim}.}
  \bibinfo{year}{2025}\natexlab{}.
\newblock \showarticletitle{Hybe: {GPU-NPU} Hybrid System for Efficient {LLM}
  Inference with Million-Token Context Window}. In
  \bibinfo{booktitle}{\emph{Proceedings of the 52nd Annual International
  Symposium on Computer Architecture}} \emph{(\bibinfo{series}{ISCA '25})}.
  \bibinfo{publisher}{ACM}, \bibinfo{address}{New York, NY, USA},
  \bibinfo{pages}{808--820}.
\newblock


\bibitem[{Nvidia}(2022)]%
        {nvidia-agx-orin-tdg}
\bibfield{author}{\bibinfo{person}{{Nvidia}}.} \bibinfo{year}{2022}\natexlab{}.
\newblock \bibinfo{booktitle}{\emph{{Jetson AGX Orin Series Thermal Design
  Guide}}}.
\newblock
\urldef\tempurl%
\url{https://www.mouser.com/pdfDocs/Jetson_AGX_Orin_Series_TDG-10943-001_v11.pdf}
\showURL{%
\tempurl}
\newblock
\shownote{Accessed: 2026-09-03}.


\bibitem[{Nvidia}(2026a)]%
        {nvidia-cutlass-gemm}
\bibfield{author}{\bibinfo{person}{{Nvidia}}.}
  \bibinfo{year}{2026}\natexlab{a}.
\newblock \bibinfo{booktitle}{\emph{{CUTLASS: Efficient GEMM in CUDA}}}.
\newblock
\urldef\tempurl%
\url{https://docs.nvidia.com/cutlass/latest/media/docs/cpp/efficient_gemm.html}
\showURL{%
\tempurl}
\newblock
\shownote{Accessed: 2026-08-10}.


\bibitem[{Nvidia}(2026b)]%
        {nvidia-jetson-orin-thermal}
\bibfield{author}{\bibinfo{person}{{Nvidia}}.}
  \bibinfo{year}{2026}\natexlab{b}.
\newblock \bibinfo{booktitle}{\emph{{Jetson Orin Platform Power and
  Performance: Thermal Management}}}.
\newblock
\urldef\tempurl%
\url{https://docs.nvidia.com/jetson/archives/r36.4.4/DeveloperGuide/SD/PlatformPowerAndPerformance/JetsonOrinNanoSeriesJetsonOrinNxSeriesAndJetsonAgxOrinSeries.html}
\showURL{%
\tempurl}
\newblock
\shownote{Accessed: 2026-06-02}.


\bibitem[{Nvidia Jetson AI Lab}(2026)]%
        {nvidia-jetson-vllm}
\bibfield{author}{\bibinfo{person}{{Nvidia Jetson AI Lab}}.}
  \bibinfo{year}{2026}\natexlab{}.
\newblock \bibinfo{title}{Serving Large Language and Vision-Language Models
  with {vLLM} on Jetson}.
\newblock
\urldef\tempurl%
\url{https://www.jetson-ai-lab.com/tutorials/genai-on-jetson-llms-vlms/}
\showURL{%
\tempurl}
\newblock
\shownote{Accessed: 2026-08-08}.


\bibitem[Raissi et~al\mbox{.}(2019)]%
        {raissi2019pinn}
\bibfield{author}{\bibinfo{person}{Maziar Raissi}, \bibinfo{person}{Paris
  Perdikaris}, {and} \bibinfo{person}{George~Em Karniadakis}.}
  \bibinfo{year}{2019}\natexlab{}.
\newblock \showarticletitle{Physics-Informed Neural Networks: A Deep Learning
  Framework for Solving Forward and Inverse Problems Involving Nonlinear
  Partial Differential Equations}.
\newblock \bibinfo{journal}{\emph{J. Comput. Phys.}}  \bibinfo{volume}{378}
  (\bibinfo{year}{2019}), \bibinfo{pages}{686--707}.
\newblock


\bibitem[{Runpod}(2026)]%
        {runpod2026stateofai}
\bibfield{author}{\bibinfo{person}{{Runpod}}.} \bibinfo{year}{2026}\natexlab{}.
\newblock \bibinfo{booktitle}{\emph{The State of {AI}}}.
\newblock \bibinfo{type}{{T}echnical {R}eport}. \bibinfo{institution}{Runpod}.
\newblock
\urldef\tempurl%
\url{https://24119306.fs1.hubspotusercontent-na1.net/hubfs/24119306/eBooks\%20and\%20Whitepapers/Runpod-State-of-AI-dark.pdf}
\showURL{%
\tempurl}


\bibitem[Sridhar et~al\mbox{.}(2010)]%
        {sridhar2010_3dice}
\bibfield{author}{\bibinfo{person}{Arvind Sridhar}, \bibinfo{person}{Alessandro
  Vincenzi}, \bibinfo{person}{Martino Ruggiero}, \bibinfo{person}{Thomas
  Brunschwiler}, {and} \bibinfo{person}{David Atienza}.}
  \bibinfo{year}{2010}\natexlab{}.
\newblock \showarticletitle{{3D-ICE}: Fast Compact Transient Thermal Modeling
  for {3D ICs} with Inter-Tier Liquid Cooling}. In
  \bibinfo{booktitle}{\emph{Proceedings of the 2010 IEEE/ACM International
  Conference on Computer-Aided Design}}. \bibinfo{pages}{463--470}.
\newblock


\bibitem[Stojkovic et~al\mbox{.}(2025)]%
        {stojkovic2025tapas}
\bibfield{author}{\bibinfo{person}{Jovan Stojkovic}, \bibinfo{person}{Chaojie
  Zhang}, \bibinfo{person}{{\'I}{\~n}igo Goiri}, \bibinfo{person}{Esha
  Choukse}, \bibinfo{person}{Haoran Qiu}, \bibinfo{person}{Rodrigo Fonseca},
  \bibinfo{person}{Josep Torrellas}, {and} \bibinfo{person}{Ricardo
  Bianchini}.} \bibinfo{year}{2025}\natexlab{}.
\newblock \showarticletitle{{TAPAS}: Thermal- and Power-Aware Scheduling for
  {LLM} Inference in Cloud Platforms}. In \bibinfo{booktitle}{\emph{Proceedings
  of the 30th ACM International Conference on Architectural Support for
  Programming Languages and Operating Systems, Volume 2}}
  \emph{(\bibinfo{series}{ASPLOS '25})}. \bibinfo{publisher}{ACM},
  \bibinfo{address}{New York, NY, USA}, \bibinfo{pages}{1266--1281}.
\newblock


\bibitem[Stuart(2010)]%
        {stuart2010inverse}
\bibfield{author}{\bibinfo{person}{Andrew~M. Stuart}.}
  \bibinfo{year}{2010}\natexlab{}.
\newblock \showarticletitle{Inverse problems: A {Bayesian} perspective}.
\newblock \bibinfo{journal}{\emph{Acta Numerica}}  \bibinfo{volume}{19}
  (\bibinfo{year}{2010}), \bibinfo{pages}{451--559}.
\newblock


\bibitem[Terraneo et~al\mbox{.}(2022)]%
        {terraneo2022_3dice}
\bibfield{author}{\bibinfo{person}{Federico Terraneo}, \bibinfo{person}{Alberto
  Leva}, \bibinfo{person}{William Fornaciari}, \bibinfo{person}{Marina
  Zapater}, {and} \bibinfo{person}{David Atienza}.}
  \bibinfo{year}{2022}\natexlab{}.
\newblock \showarticletitle{{3D-ICE} 3.0: Efficient Nonlinear {MPSoC} Thermal
  Simulation With Pluggable Heat Sink Models}.
\newblock \bibinfo{journal}{\emph{IEEE Transactions on Computer-Aided Design of
  Integrated Circuits and Systems}} \bibinfo{volume}{41}, \bibinfo{number}{4}
  (\bibinfo{year}{2022}), \bibinfo{pages}{1062--1075}.
\newblock


\bibitem[{U.S. Environmental Protection Agency}(2025)]%
        {epa-extreme-heat}
\bibfield{author}{\bibinfo{person}{{U.S. Environmental Protection Agency}}.}
  \bibinfo{year}{2025}\natexlab{}.
\newblock \bibinfo{title}{{Extreme Heat and Climate Change}}.
\newblock
\urldef\tempurl%
\url{https://www.epa.gov/climatechange-science/extreme-heat}
\showURL{%
\tempurl}
\newblock
\shownote{Accessed: 2026-06-02}.


\bibitem[{vLLM Development Team}(2026)]%
        {vllm-v1-guide}
\bibfield{author}{\bibinfo{person}{{vLLM Development Team}}.}
  \bibinfo{year}{2026}\natexlab{}.
\newblock \bibinfo{title}{{vLLM V1 User Guide}}.
\newblock
\urldef\tempurl%
\url{https://github.com/vllm-project/vllm/blob/main/docs/usage/v1_guide.md}
\showURL{%
\tempurl}
\newblock
\shownote{Accessed: 2026-08-08}.


\bibitem[Xu et~al\mbox{.}(2025)]%
        {xu2025fastnpu}
\bibfield{author}{\bibinfo{person}{Daliang Xu}, \bibinfo{person}{Hao Zhang},
  \bibinfo{person}{Liming Yang}, \bibinfo{person}{Ruiqi Liu},
  \bibinfo{person}{Gang Huang}, \bibinfo{person}{Mengwei Xu}, {and}
  \bibinfo{person}{Xuanzhe Liu}.} \bibinfo{year}{2025}\natexlab{}.
\newblock \showarticletitle{Fast On-Device {LLM} Inference with {NPUs}}. In
  \bibinfo{booktitle}{\emph{Proceedings of the 30th ACM International
  Conference on Architectural Support for Programming Languages and Operating
  Systems, Volume 1}} \emph{(\bibinfo{series}{ASPLOS '25})}.
  \bibinfo{publisher}{ACM}, \bibinfo{address}{New York, NY, USA},
  \bibinfo{pages}{445--462}.
\newblock


\bibitem[Yang et~al\mbox{.}(2021)]%
        {yang2021bpinns}
\bibfield{author}{\bibinfo{person}{Liu Yang}, \bibinfo{person}{Xuhui Meng},
  {and} \bibinfo{person}{George~Em Karniadakis}.}
  \bibinfo{year}{2021}\natexlab{}.
\newblock \showarticletitle{{B-PINNs}: {Bayesian} physics-informed neural
  networks for forward and inverse {PDE} problems with noisy data}.
\newblock \bibinfo{journal}{\emph{J. Comput. Phys.}}  \bibinfo{volume}{425}
  (\bibinfo{year}{2021}), \bibinfo{pages}{109913}.
\newblock


\bibitem[Zhang et~al\mbox{.}(2026)]%
        {zhang2026core}
\bibfield{author}{\bibinfo{person}{Zongpu Zhang}, \bibinfo{person}{Pranab
  Dash}, \bibinfo{person}{Qiang Xu}, \bibinfo{person}{Y.~Charlie Hu},
  \bibinfo{person}{Jian Li}, {and} \bibinfo{person}{Haibing Guan}.}
  \bibinfo{year}{2026}\natexlab{}.
\newblock \showarticletitle{Rethinking {DVFS} for Mobile {LLM}s: Unified
  Energy-Aware Scheduling with {CORE}}. In
  \bibinfo{booktitle}{\emph{Proceedings of Machine Learning and Systems (MLSys
  2026)}}, Vol.~\bibinfo{volume}{8}. \bibinfo{pages}{186--198}.
\newblock


\bibitem[Zhou et~al\mbox{.}(2025)]%
        {zhou2025asrrpinn}
\bibfield{author}{\bibinfo{person}{Zijian Zhou}, \bibinfo{person}{Min Tang},
  {and} \bibinfo{person}{Liang Chen}.} \bibinfo{year}{2025}\natexlab{}.
\newblock \showarticletitle{{ASRR-PINN}: Adaptive Sub-Regional Random
  Resampling-Based {PINN} for Thermal Analysis of {3D-ICs}}. In
  \bibinfo{booktitle}{\emph{2025 62nd ACM/IEEE Design Automation Conference
  (DAC)}}. \bibinfo{publisher}{IEEE}, \bibinfo{address}{Piscataway, NJ, USA},
  \bibinfo{pages}{1--7}.
\newblock


\end{thebibliography}

\end{document}